\documentclass[11pt]{article}
\usepackage[utf8]{inputenc}
\usepackage{authblk}
\usepackage{natbib}
\usepackage{graphicx}
\usepackage{amsmath}
\usepackage{bm}
\usepackage{setspace}
\usepackage{fullpage} 
\usepackage{natbib}
\usepackage{bbm}
\usepackage{subcaption}
\usepackage{hyperref}
\hypersetup{
    colorlinks=true,
    linkcolor=blue,
    filecolor=magenta,      
    urlcolor=cyan,
    citecolor=blue
}

\usepackage{adjustbox}              
\usepackage{caption}                
\usepackage{rotating}               
    \numberwithin{table}{section}   
\usepackage{booktabs}               
\usepackage{dcolumn}   
\usepackage{tikz}
    \newcolumntype{d}[1]{D{.}{.}{#1}}

\newcommand{\sym}[1]{\rlap{#1}} 

\title{High-frequency lead-lag relationships in the Chinese stock index futures market: tick-by-tick dynamics of calendar spreads}
\author[1]{Guanlin Li \thanks{Email: guanlinl131205@163.com}}
\author[2]{Xiyan Chen \thanks{Email: chenxiyan@amss.ac.cn}}
\affil[1,2]{China Securities Co., Ltd., Beijing, China}
\affil[2]{Academy of Mathematics and Systems Science, Chinese Academy of Sciences, Beijing, China}
\author[3]{Yingzheng Liu \thanks{Email: yliu2@trinity.edu}}
\affil[3]{Finance and Business Analytics, Trinity University, San Antonio, Texas, United States}

\begin{document}
\maketitle

\begin{abstract}
Lead-lag relationships, integral to market dynamics, offer valuable insights into the trading behavior of high-frequency traders (HFTs) and the flow of information at a granular level. This paper investigates the lead-lag relationships between stock index futures contracts of different maturities in the Chinese financial futures market (CFFEX). Using high-frequency (tick-by-tick) data, we analyze how price movements in near-month futures contracts influence those in longer-dated contracts, such as next-month, quarterly, and semi-annual contracts. Our findings reveal a consistent pattern of price discovery, with the near-month contract leading the others by one tick, driven primarily by liquidity. Additionally, we identify a negative feedback effect of the "lead-lag spread" on the leading asset, which can predict returns of leading asset. Backtesting results demonstrate the profitability of trading based on the lead-lag spread signal, even after accounting for transaction costs. Altogether, our analysis offers valuable insights to understand and capitalize on the evolving dynamics of futures markets.
\end{abstract}

\section{Introduction}\label{sect: intro}

The lead-lag effect in financial markets refers to a delayed relationship in which the price movement of one asset influences the future price of another, often with a time gap. This phenomenon is especially significant in high-frequency trading (HFT), where sophisticated algorithms and powerful computers detect and exploit these small time delays to make rapid profits (\citealp{chaboud2014rise, budish2015high, biais2015equilibrium}). While these lead-lag effects create opportunities for statistical arbitrage, they are typically short-lived because market participants quickly recognize and act on these patterns, thereby correcting any pricing inefficiencies (\citealp{huth2014high, alsayed2014ultra, dao2018ultra}). Financial markets are highly complex, with numerous interconnected assets and instruments that influence each other in various ways. Understanding the pairwise relationships between these assets is crucial for analyzing market dynamics at a granular level. By identifying and accurately predicting lead-lag patterns, traders can anticipate price movements, manage risk, and make more informed decisions. However, fully capitalizing on these opportunities can be hindered by market frictions, such as transaction costs, liquidity constraints, and other inefficiencies (\citealp{alsayed2014ultra, huth2014high}). The lead-lag effect has been explored across various financial markets and products, including foreign exchange (\citealp{mizuno2006correlation, basnarkov2020lead}), stock prices (\citealp{li2021dynamic}), stock indexes and their corresponding futures contracts (\citealp{kawaller1987temporal, frino2000lead, gwilym2001lead, kavussanos2008lead}).

A substantial body of research focuses on the lead-lag relationship between stock index markets and futures markets. For example, in the Chinese market, the relationship between the CSI 300 index and its corresponding futures contract (IF) has been studied (\citealp{hou2013price, hou2014impact, wang2017lead, ma2022measuring}). Previous studies have shown that, in high-frequency data, the futures contract leads the spot index. This can be attributed to the fact that futures contracts are traded intraday, while the spot index and underlying stocks are not. Moreover, the futures market offers a higher degree of leverage. Our study aims to investigate how information flows across futures contracts with different maturities on a tick-by-tick basis. The standard stock index futures on the China Financial Futures Exchange (CFFEX), such as IF, have four maturities: near-month, next-month, quarterly, and semi-annual. To the best of our knowledge, the tick-by-tick lead-lag relationship between futures contracts of different maturities has not been thoroughly examined.

Correlation-based methods are widely used to identify lead-lag relationships in high-frequency data. If the past returns of asset $X$ can predict the future returns of asset $Y$, we say that $X$ leads $Y$. These methods include cross-correlation analysis, Granger causality tests, and others (\citealp{granger1969investigating, hayashi2005covariance, huth2012some, hoffmann2013estimation}). We employ the Hayashi–Yoshida (HY) cross-correlation estimator (\citealp{hayashi2005covariance}) to measure lead-lag relationships. This estimator addresses asynchronous trading without relying on sampling, utilizing all available tick-by-tick data, which are inherently irregularly sampled. Another approach to examine lead-lag relationships involves identifying the optimal causal path between two time series, aiming to minimize the mismatch between them. Although this provides dynamic lead-lag estimates, it is sensitive to noise. In literature, a robust version of this approach that applies the Boltzmann factor to assign probabilities to different paths. However, determining the appropriate level of averaging (the "temperature" in the Boltzmann factor) remains a challenge (\citealp{sornette2005non, wang2017lead, stubinger2019statistical}). In this study, we apply a simple block bootstrap method (\citealp{bollen2017tail, bangsgaard2024lead}) to test the significance of the lead-lag relationship and examine its volatility over time.

After performing pairwise lead-lag analyses and statistical tests for each pair of futures contracts with different maturities, we identify a total of 6 pairs of futures contracts (choose two from four maturities) for each index future. Our results show that the most liquid futures contract (typically the one with the nearest expiration date and the highest trading volume) consistently leads the other three contracts (next-month, quarterly, and semi-annual contracts). We refer to this contract as the "lead future." These findings are consistent with existing research, which suggests that more liquid assets tend to lead less liquid assets (\citealp{de1997high, jong1998intraday, lo1990econometric, jarnecic1999trading, kadlec1999transactions, hou2001information, toth2006increasing, huth2014high}). Furthermore, our analysis of joint movements (e.g. principle component analysis, PCA) reveals that a robust principal component explains the tick-by-tick variations across futures contracts with different maturities. Specifically, when the lead future moves by one tick, the other three contracts tend to follow, each moving by approximately one tick as well, indicating a level shift across maturities. However, other decomposed structural movements resulted from PCA are less consistent and tend to vary on a daily basis.

Following the identification of significant lead-lag relationships, many studies attempt to backtest strategies that could be profitable by predicting the returns of lagging assets based on observed leading assets. While lead-lag relationships may suggest potential arbitrage opportunities, market frictions such as bid-ask spreads and transaction costs can make these opportunities difficult to exploit profitably. Some advanced econometric models have been shown to work even after accounting for these costs (\citealp{stubinger2019statistical, poutre2024profitability}). Instead of focusing on arbitrage strategies based on a finely tuned forecasting model that predicts the returns of lagging assets from the dynamics of leading assets, our study aims to leverage the dynamic relationship between leading and lagging futures contracts to predict the returns of the leading asset itself. We hypothesize that high-frequency traders (HFTs) could profit by trading the "lead-lag spread" when it deviates significantly, indicating a weakening of the relationship (during volatile market conditions). We suggest that this "spread" or weakening of the relationship may signal a negative feedback effect on the lead future. When the spread widens, for example, when the leading asset rises much faster than the lagging asset, HFTs may sell the lead asset and buy the lagging asset, inhibiting the momentum of the leading asset. Thus, the short-term lead-lag spread could influence the return of the leading asset, potentially creating a profitable trading signal. To the best of our knowledge, this is the first study to explicitly use the spread between leading and lagging assets to predict the return of the leading asset on a tick-by-tick basis.

Lastly, we propose a simple trading strategy based on the negative feedback effect of the lead-lag spread on the return of the leading asset. This strategy proves to be profitable on out-of-sample data, even after accounting for market frictions. The strategy performs exceptionally well during periods of high market volatility, which aligns with the characteristics of an 
(statistical) arbitrage strategy (\citealp{marshall2013etf}). The trading behavior of this strategy resembles a "contrarian" approach, as it takes positions that counteract market overreactions. In this paper, we rigorously examine that the predictive power of the "lead-lag spread" primarily arises from the cross-effect between the leading and lagging assets, rather than from any negative autocorrelation within either the leading or lagging asset. This finding offers a microscopic perspective on the cross-effects of market dynamics, as discussed in \citealp{lo1990contrarian}. In contrast to that study, our research shows that the overreaction to the "lead-lag spread" could be profitable for the leading asset.

The remainder of the paper is structured as follows. Section \ref{sect: data} introduces the high-frequency data structure in CFFEX and how we process the raw data, as well as the basic liquidity measures of different futures contracts. Section \ref{sect: method} presents the statistical estimators and models used to analyze tick by tick lead-lag networks. It also introduces the statistical testing procedure used to validate the lead-lag spreads feedback effect on lead futures contract. Section \ref{sect: results} presents the results and their implications. Finally, we conclude the paper in Section \ref{sect: conclusion}.

\section{Data description and summary statistics}\label{sect: data}

\subsection{Stock index futures in CFFEX}
Stock index futures contracts are traded on China Financial Futures Exchange (CFFEX), such as IF, IC, IM, and IH, track different indices that represent varying market segments, from large-cap to small-cap stocks. IF tracks the CSI 300 Index, covering the top 300 large-cap stocks from both Shanghai and Shenzhen, with high liquidity and moderate volatility. IC follows the CSI 500 Index, focusing on 500 mid-cap stocks, offering higher growth potential but with more volatility. IM tracks the CSI 1000 Index, which includes the smallest 1000 stocks, providing the highest growth potential at the cost of higher risk and lower liquidity. IH tracks the SSE 50 Index, representing the top 50 large-cap stocks on the Shanghai Stock Exchange, offering stability and liquidity but narrower market coverage. The main differences lie in the composition, liquidity, and risk profile of the contracts. Investors choose based on their desired exposure and risk tolerance. The index futures has four maturities, near-month, next-month, quarterly, and semi-annual.

\subsection{Reformatting data into time-tick structure}
Our dataset includes all four types of futures contracts (IF, IH, IC, and IM) traded on the CFFEX between January 1, 2022, and November 25, 2024, covering approximately 700 trading days. We access real-time limit order book (LOB) data via a server co-located near the CFFEX. However, detailed information on trades, quotes, or canceled orders is not available to market participants. The LOB data represents the most granular level of information accessible for our analysis. Our analysis primarily focuses on Level 1 data, which includes bid and ask prices, volumes, spreads, and other related metrics. 

\begin{figure}[!ht]
\centering
\includegraphics[width=0.9\textwidth]{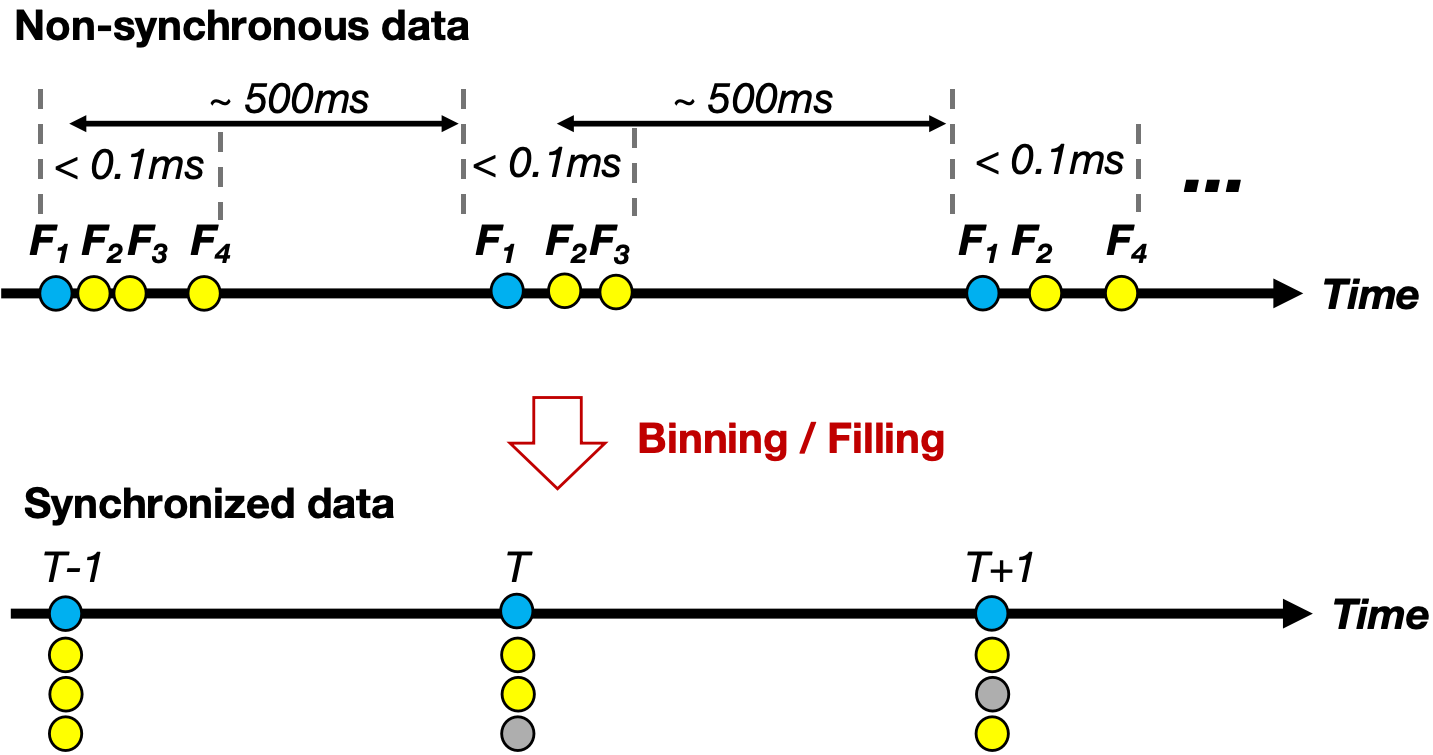}
\caption{Schematic diagram of transforming non-synchronized raw data to synchronized data. This diagram illustrates the process of converting non-synchronized raw data into synchronized data. There are three LOB events: $T-1$, $T$, and $T+1$. (Top) At the first event, $T-1$, futures contracts across four maturities are updated. However, the update for $F_{4}$ is missing at time $T$, and the update for $F_{3}$ is missing at time $T+1$. For the three observed LOB update events, the time between the arrival of the first contract (e.g., $F_{1}$) and the last arrived contract (e.g., $F_{4}$) is less than $0.1$ milliseconds. (Bottom) After the binning and filling procedures, the missing data is imputed, and the update times for different contracts within a LOB event are adjusted to align with the time of the first arriving contract. Note that the times, $T-1,\ T$ and $T+1$, correspond to points on the tick grid.}
\label{fig: sychdata}
\end{figure}

For the IF (CSI 300 index) futures contracts, as well as the IH, IC, and IM contracts, we receive limit order book (LOB) update events approximately every 500 milliseconds. Every update include changes to the order books for all maturities of the IF contracts. However, if there are no changes in the order book, no new order book is transmitted. The LOBs for different maturities in a single update event are not received simultaneously; instead, they arrive in quick succession and are sorted alphabetically by instrument name. Within a single update event, the contracts are received in the following order: near-month, next-month, quarterly, and semi-annual. Summary statistics of a sample data, including the time gap between the first and last contracts in an update event, as well as the time gap between consecutive update events, are provided in Table \ref{tab: updateTimingsInLocalServer}. The market data update process is shown in the top panel in Fig.~\ref{fig: sychdata}.

\begin{table}[!htbp]
\caption{Summary statistics comparing the time gaps between consecutive update events and the duration of each update event} \label{tab: updateTimingsInLocalServer}
\centering
    {
\def\sym#1{\ifmmode^{#1}\else\(^{#1}\)\fi}
\begin{tabular}{l*{1}{ccccc}}
\toprule
                    &         Mean&        $Q_{.05}$&          Median&         $Q_{.95}$&        N\\
\midrule
Time interval between consecutive events          &      500.01&     495.98&     500.11&         504.56&      28914\\
Duration of an event   &      0.03&      0.01&      0.03&          0.05&       28915\\
\bottomrule
\end{tabular}
}
    
\small{Note: Using the last trading day in our dataset (e.g., November 25, 2024) as sample data, we show the characteristics of the timing of the limit orderbook data stream for IF. The statistics are measured in milliseconds (ms). With 4 trading hours and updates every 500ms, we collect about 28,000 samples.}
\end{table}

\vfill

We transform our non-synchronized LOB data streams (futures contract with different maturities) into synchronized tick timestamps through two steps: binning and filling. First, we combine all updates from a single event into one update, using the timestamp of the first futures contract received as the timestamp for the combined event. The 95th percentile ($Q_{.95}$) of the time gap between receiving the first and last futures contracts in an event is approximately 0.05 ms (see Table \ref{tab: updateTimingsInLocalServer}), which indicates that there is insufficient time for market participants to pass on information from the first contract to the last within the event. This binning process effectively shifts the longer-tenor contracts (next-month, quarterly, and semi-annual) forward in time, while relatively delaying the near-month contracts. As a result, we use the transformed data where near-month updates are delayed to emphasize the leading effect of near-month futures, thus strengthening the lead-lag relationships.

Next, we fill in any missing futures contracts in an update event by using the data from the previous tick. While we recognize that filling missing values via previous-tick interpolation can introduce spurious lead-lag effects, the correlation level may decrease when asset processes are sampled synchronously at high frequencies (\citealp{epps1979comovements}). In our case, however, the futures contracts across different maturities are observed synchronously. If a futures contract update is missing in an event, it simply means that the contract’s order book is the same as the previous tick, with no statistical uncertainty or assumptions required. In Section \ref{sect: results}, we will demonstrate that there is no significant difference between the lead-lag analysis results of the raw and transformed data. The futures contracts in the transformed data have the same tick timestamps, which makes tick-by-tick analysis of contracts with different maturities more convenient. Instead of working with absolute time (e.g., in seconds or milliseconds), this approach allows us to shift the process forward or backward by units of ticks, rather than by a fixed time duration. 

\subsection{Summary statistics}
As previously mentioned, each futures class consists of four maturities, typically identified by their contract codes, for example of IF, e.g., IF2407 for near-month, IF2408 for next-month, IF2409 for quarterly, and IF2412 for semi-annual. The near-month contracts generally have the highest trading volume, although this can shift as the near-month contract approaches expiration, with the next-month contract taking its place. For simplicity, we classify these contracts into four liquidity-based names, $F_{1}$, $F_{2}$, $F_{3}$, and $F_{4}$, ranked by trading volume. While these code assignments may vary on a daily basis depending on market conditions, $F_{1}$ typically corresponds to the near-month contract, $F_{2}$ to the next-month contract, $F_{3}$ to the quarterly contract, and $F_{4}$ to the semi-annual contract.

Table \ref{tab: avg of F1 and F2's stats} shows the average values of trading volume, bid-ask spread, tick change, and annualized volatility for all stock index futures in CFFEX. For each stock index future (e.g., IF), the more liquid futures contract $F_{1}$ typically exhibits higher daily trading volume and a tighter bid-ask spread. 
The annualized volatility remains approximately the same across different maturities. Note that $F_{3}$ and $F_{4}$ are not included in the table due to space constraints, but their results are consistent with $F_{1}$ and $F_{2}$, as they are even less liquid. When comparing different stock index futures, we observe that the volatility for IM and IC are higher than for IF and IH. This difference arises from the different underlying components of these futures contracts, as detailed in Section \ref{sect: data}.

\begin{table}[!htbp]
    \caption{Average values for trading volume, bid-ask spread and annualized volatility for all stock index futures in CFFEX, across two maturities: $F_{1}$ and $F_{2}$} \label{tab: avg of F1 and F2's stats}
    \centering
    {
\def\sym#1{\ifmmode^{#1}\else\(^{#1}\)\fi}
\begin{tabular}{l*{8}{D{.}{.}{-1}}}
\toprule
                    &\multicolumn{2}{c}{IC}              &\multicolumn{2}{c}{IF}           &\multicolumn{2}{c}{IH}  &\multicolumn{2}{c}{IM}\\\cmidrule(lr){2-3}\cmidrule(lr){4-5}\cmidrule(lr){6-7}\cmidrule(lr){8-9}
                    &\multicolumn{1}{c}{$F_{1}$}         &\multicolumn{1}{c}{$F_{2}$}      &\multicolumn{1}{c}{$F_{1}$}         &\multicolumn{1}{c}{$F_{2}$}         &\multicolumn{1}{c}{$F_{1}$}         &\multicolumn{1}{c}{$F_{2}$}          &\multicolumn{1}{c}{$F_{1}$}         &\multicolumn{1}{c}{$F_{2}$}\\
\midrule
Trading volume                 &       4.82&       1.98&      6.02&      2.19&       3.50&       1.38&       4.24&      1.78\\
\addlinespace
Level-1 bid-ask spread             &      0.75&      1.15&       0.43&       0.61&      0.38&      0.50&       0.88&       1.23\\
\addlinespace
Volatility             &      0.15&      0.15&       0.13&       0.13&      0.13&      0.13&       0.16&       0.16\\
\bottomrule
\end{tabular}
}
        
    \small{Note: The table presents the average values of market metrics for all trading days. Trading volume is measured in units of 10,000 lots. The level-1 bid-ask spread is measured in price terms. To calculate it, we first average the bid-ask spread across all LOBs intraday, then compute the average across all trading days. The mid-quote change is calculated in the same way as the bid-ask spread and tracks the tick-by-tick changes in asset prices. To compute realized variance for a day, we sum the 5-minute return variances and multiply the result by 252 to obtain the annualized realized variance. The annualized volatility is then calculated as the square root of the annualized variance.} 
\end{table}

\section{Methods}\label{sect: method}

\subsection{lead-lag estimators}
We work with high-frequency data, which are often characterized by two main challenges: non-synchronicity and irregular observation intervals. To address these issues, 
\citealp{hayashi2005covariance,hoffmann2013estimation} propose a covariance estimator of non-synchronously observed diffusion processes. Consider two observed processes, $\{X_{t}\}$ and $\{Y_{t}\}$, the discrete observation times are, $t_{1}^{X} < t_{2}^{X} < ... < t_{n}^{X}$ and $t_{1}^{Y} < t_{2}^{Y} < ... < t_{n}^{Y}$, the $i$th observed interval of $\{X_{t}\}$-process is $I_{i}^{X} = (t_{i-1}^{X}, t_{i}^{X}]$, the $j$th observed interval of $\{Y_{t}\}$-process is $I_{j}^{Y} = (t_{j-1}^{Y}, t_{j}^{Y}]$. The time intervals, $I_{i}^{X}$ and $I_{j}^{Y}$ are considered overlapped when the observation periods of the two processes coincide or partially coincide, we use indicator function, $\chi_{i,j}=1$ if $I_{i}^{X} \cap I_{j}^{Y} \neq  \emptyset$, otherwise $\chi_{i,j}=0$. The Hayashi–Yoshida (HY) cross-correlation estimator of two observed processes, $\{X_{t}\}$ and $\{Y_{t}\}$, is
\begin{align}\label{eq: HY-estimator}
\begin{split}
\hat{\rho}_{HY} = \frac{\sum_{i}\sum_{j} (\Delta X_{i} \Delta Y_{j}) \chi_{i,j}}{\sqrt{\sum_{i}{(\Delta X_{i})^{2}}\sum_{j}{(\Delta Y_{j})^{2}}}},
\end{split}
\end{align}	
where $\Delta X_{i} = X(t_{i}^{X})-X(t_{i-1}^{X})$ and $\Delta Y_{j} = Y(t_{j}^{Y})-Y(t_{j-1}^{Y})$. In this paper, we will primarily use the transformed data (see Section \ref{sect: data} for binning and filling procedures), the cross-correlation estimator $\hat{\rho}_{HY}$ for synchronized and regularly sampled observations simplifies to the standard cross-correlation estimator. To explicitly estimate the timing relationship between two processes, we introduce the lag parameter $\ell$, the $j$th interval of shifted process $Y_{t}^{(\ell)}$  is $(t_{j-1}^{Y} + \ell, t_{j}^{Y} + \ell]$, the cross-correlation estimator of process $X_{t}$ and shifted process $Y_{t}^{(\ell)}$ is denoted by $\hat{\rho}_{HY}(\ell)$.

We examine three measures for assessing the lead-lag relationship between two instruments: Lead-Lag Time (LLT), Lead-Lag Correlation (LLC), and Lead-Lag Ratio (LLR) (\citealp{huth2014high}). LLT identifies the time at which the absolute value of the cross-correlation function is maximized, indicating the lead-lag direction between the instruments. LLC measures the corresponding value of the cross-correlation at this point. In contrast, LLR summarizes the entire cross-correlation function into a single value that reflects the asymmetry of the relationship. LLR is defined as 
\begin{align}\label{eq: LLR}
\begin{split}
\text{LLR} = \frac{\sum_{i=1}^{n} \hat{\rho}_{HY}(\ell_{i})^{2}}{\sum_{i=1}^{n} \hat{\rho}_{HY}(-\ell_{i})^{2}},
\end{split}
\end{align}	
where $\ell_{1}, ..., \ell_{n}$ are discrete positive time lags. While LLT is sensitive to the shape of the cross-correlation, LLR provides a more comprehensive measure of the overall strength of the lead-lag relationship. When $LLR > 1$, it indicates that process $X_{t}$ leads $Y_{t}$; otherwise, $Y_{t}$ leads $X_{t}$. Together, these measures complement each other: LLT focuses on specific time points, while LLR captures the broader structure of the relationship. Notably, LLR may lead to a different conclusion about the lead-lag relationship compared to LLT (\citealp{bangsgaard2024lead}).

To test the statistical significance of the intraday lead-lag relationship, we apply a block bootstrap technique (\citealp{bollen2017tail, bangsgaard2024lead}). This method resamples the data in blocks, rather than individual data points, to preserve the time-series dependencies. The choice of block size is crucial: too small a block may lose temporal dependencies, while too large a block reduces variability. We divide each trading day into 24 intervals of 10 minutes and resample them 2,000 times to generate an empirical distribution of lead-lag measures (e.g., LLR and LLT).

In Section \ref{sect: results}, we demonstrate that the most liquid contract, typically the near-month futures, leads the other contracts by one tick. This relationship holds consistently across all CFFEX stock index futures (IF, IH, IC, and IM) over the approximately 700 trading days covered in this study. There are four maturities for a stock future, which allows us to examine the lead-lag relationships for six pairs of futures contracts (i.e., the number of ways to choose 2 contracts from 4). We then extract the characteristic lead-lag tree by finding the minimum spanning tree (MST) from the pairwise lead-lag networks (\citealp{mantegna1999hierarchical, onnela2002dynamic}). The LLC, $\rho^{*}$, is used as the correlation measure between two instruments, and the pairwise distance is defined as $d=\sqrt{1-|\rho^{*}|}$, such that $0\leq d \leq 1$. The MST is a connected, acyclic graph (i.e., a tree) that minimizes the sum of edge distances. It extracts the most significant lead-lag correlations from all pairwise correlations, making it useful for identifying lead-lag arbitrage opportunities in a large correlation network.

\subsection{Principal component analysis of joint movements}
While we have established that the lead future consistently leads other contracts by one tick, we have not yet explored the joint dynamics between these contracts. To address this, we conduct principal component analysis (PCA) on the tick-by-tick variations of futures contracts with different maturities (\citealp{abdi2010principal}). Given that we have already observed the one-tick lead of the lead future, we shift the lead future's tick backward by one tick to align the dynamics and eliminate any mismatches caused by the lead-lag relationship. The tick-by-tick variations for the futures contracts $F_{1}$, $F_{2}$, $F_{3}$, and $F_{4}$ are as follows
\begin{align}\label{eq: tick-tick-chg}
\begin{split}
\Delta F_{t} = [\Delta F_{1}(t-1),\ \Delta F_{2}(t),\ \Delta F_{3}(t),\ \Delta F_{4}(t)],
\end{split}
\end{align}	
where $\Delta F_{i}(t) = F_{i}(t) - F_{i}(t-1)$, the time $t$ is measured in tick intervals. In this paper, we focus on the first principal component, as we found that the other components vary on a daily basis, resulting in wide pointwise confidence intervals over the trading period.

\subsection{Statistical significance of a calendar spread signal}
The lead-lag relationships between a leading futures contract and other futures contracts are significant and robust. We hypothesize that High-Frequency Traders (HFTs) will actively trade a pair, comprising the leader and the lagger, when there is a short-term, significant deviation between the two. We aim to investigate whether the spread between a leading and a lagging futures contract (referred to as the calendar spread) can predict changes in the midquote of the lead futures contract. The change in the lead futures contract’s midquote after $h$ steps (time ticks) is
\begin{align}\label{eq: h-step ret}
\begin{split}
\Delta^{(h)} F_{1} = F_{1}(t+h) - F_{1}(t),
\end{split}
\end{align}	
where $t$ is the current time. We use $F_{2}$ denotes the lagger, define the short-term spread deviation between the leader and the lagger as 
\begin{align}\label{eq: spread trending on calendar spread}
\begin{split}
\theta(t) = (F_{1}(t) - F_{2}(t)) - \phi_{C}\{F_{1}(t) - F_{2}(t)\},
\end{split}
\end{align}
where $\phi_{C}\{.\}$ refers to the Exponential Moving Average (\citealp{hunter1986exponentially}), $C$ is the cycle parameter determining the smoothing (update weight is $2/(C+1)$). Additionally, we define the short-term trend or momentum of the lead futures contract, denoted $m_{1}(t)$, is given by 
\begin{align}\label{eq: spread trending on fut}
\begin{split}
m_{1}(t) = F_{1}(t) - \phi_{C}\{F_{1}(t)\}.
\end{split}
\end{align}
The short-term trend of $F_{2}$ is $m_{2}(t)$. Note that the cycle parameter in the EMA operation can influence the goodness of model fit, but optimizing it (e.g., feature engineering) is not the focus of this study. We fix $C=50$ in this study. In this study, we aim to demonstrate that the short-term trend of the calendar spread has a negative feedback effect on the lead futures contract. We propose two statistical models to predict the lead futures contract’s return. The first model (Model1) uses only the calendar spread trend as the predictor,
\begin{align}\label{eq: model1}
\begin{split}
\Delta^{(h)} F_{1}(t) = \beta_{0} + \beta_{\theta} \theta(t) + \epsilon_{t},
\end{split}
\end{align}
where $\beta_{0}$ is the intercept and $\beta_{m}$ is the coefficient for calendar spread trend, $\epsilon_{t}$ is the residual error. The second model (Model2) introduces an additional predictor, the short-term trend of the lead futures contract, alongside the calendar spread,
\begin{align}\label{eq: model2}
\begin{split}
\Delta^{(h)} F_{1}(t) = \beta_{0} + \beta_{\theta} \theta(t) + \beta_{1} m_{1}(t) + \epsilon_{t}.
\end{split}
\end{align}
It is important to note that Model2 can be represented as a weighted combination of the short-term momentum of both the lead and the lagging futures contracts,
\begin{align}\label{eq: model2 equiv}
\begin{split}
\beta_{\theta} \theta(t) + \beta_{1} m_{1}(t) &= \beta_{\theta} ((F_{1}(t) - F_{2}(t)) - \phi_{C}\{F_{1}(t) - F_{2}(t)\}) + \beta_{1} (F_{1}(t) -\phi_{C}\{F_{1}(t)\})\\
&= \beta_{\theta} (F_{1}(t) - \phi_{C}\{F_{1}(t)\}) - \beta_{\theta} (F_{2}(t) - \phi_{C}\{F_{2}(t)\}) + \beta_{1} (F_{1}(t) - \phi_{C}\{F_{1}(t)\})\\
&= (\beta_{\theta} + \beta_{1}) (F_{1}(t) - \phi_{C}\{F_{1}(t)\}) - \beta_{\theta} (F_{2}(t) - \phi_{C}\{F_{2}(t)\})\\
&= \widetilde{\beta}_{1} m_{1}(t) + \widetilde{\beta}_{2} m_{2}(t),
\end{split}
\end{align}
where $\widetilde{\beta}_{1} = \beta_{\theta} + \beta_{1}$ and $\widetilde{\beta}_{2} = -\beta_{\theta}$. 
We test whether the short-term trend of the calendar spread, $\theta(t)$, has a negative feedback effect on the lead futures contract. Additionally, we seek to validate that this effect is not driven by mean reversion in the individual futures contracts. To do so, we will compare the fit of Model1 and Model2. If the negative feedback from the calendar spread is the primary predictor of the leader's future returns, we should observe a significant $\beta_{\theta}$ in Model1, and the inclusion of $m_{1}(t)$ and $m_{2}(t)$ in Model2 should not significantly improve the model's explanatory power, meaning $\beta_{1}$ should be statistically insignificant. However, if the momentum of either contract is important, then $\beta_{1}$ in Model2 should be statistically significant, and Model2 should yield a significantly higher R-squared than Model1. This would suggest that the short-term momentum of either (or both) the lead or lag futures is the underlying predictor of future price changes in the lead futures contract.

\section{Results}\label{sect: results}
In this section, we present the results from our statistical analysis and backtesting. In Section \ref{sec: lead-lag network}, we examine the lead-lag relationships between stock index futures with different maturities. Section \ref{sec: term structure} analyzes the tick-by-tick joint movements across these futures contracts. In Section \ref{sec: regr results}, we show how the strong linkages between futures contracts of different maturities can be used to predict the returns of the lead futures contract. Finally, in Section \ref{sec: backtest}, we evaluate the performance of a simple trading strategy based on the calendar spread signal.

\subsection{Lead-lag relationships}\label{sec: lead-lag network}
\begin{figure}[!ht]
      \centering
	   \begin{subfigure}{0.3\linewidth}
		\includegraphics[width=\linewidth]{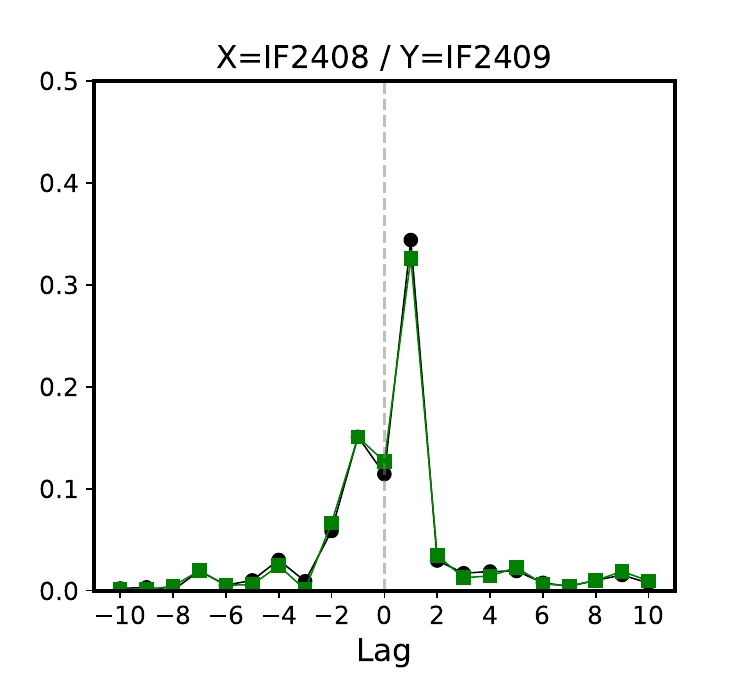}
	   \end{subfigure}
	   \begin{subfigure}{0.3\linewidth}
		\includegraphics[width=\linewidth]{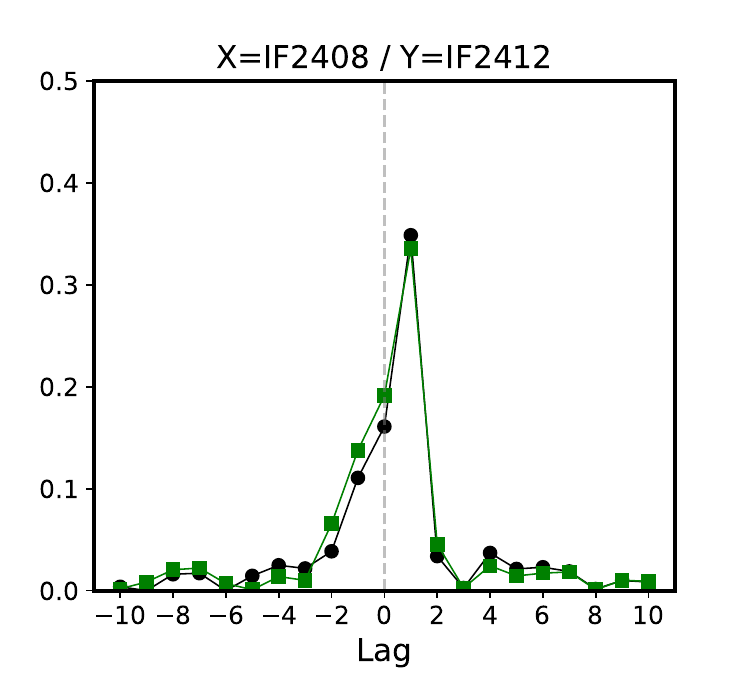}
	    \end{subfigure}
	     \begin{subfigure}{0.3\linewidth}
		 \includegraphics[width=\linewidth]{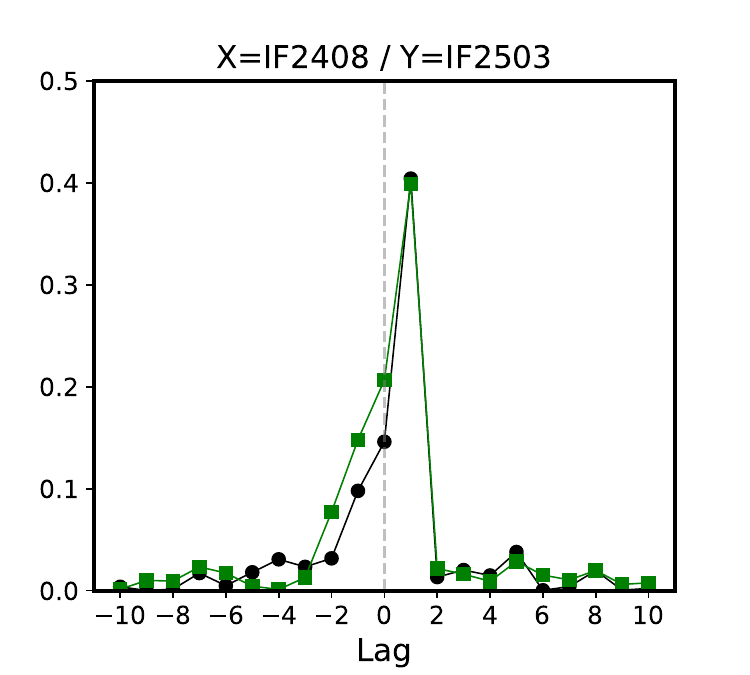}
	      \end{subfigure}
	\vfill
	       \begin{subfigure}{0.3\linewidth}
		  \includegraphics[width=\linewidth]{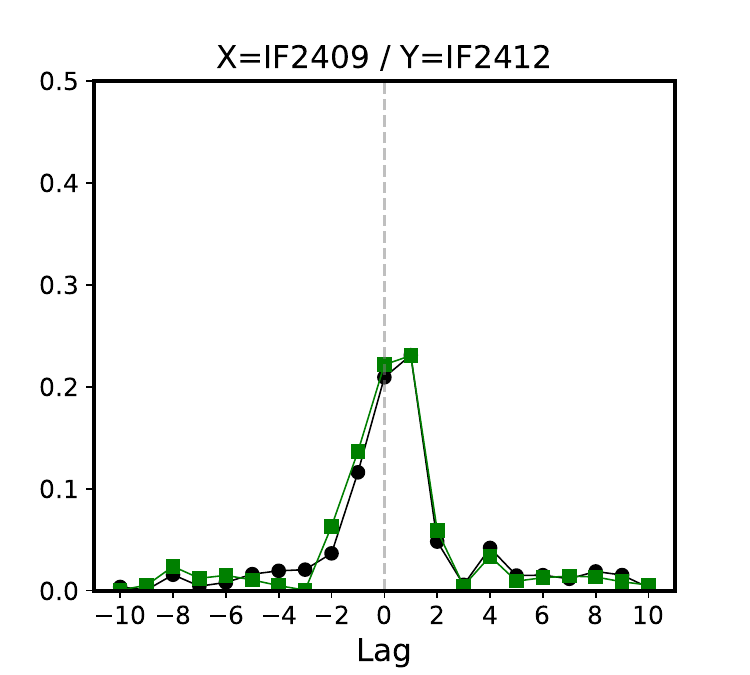}
	       \end{subfigure}
	       \begin{subfigure}{0.3\linewidth}
		  \includegraphics[width=\linewidth]{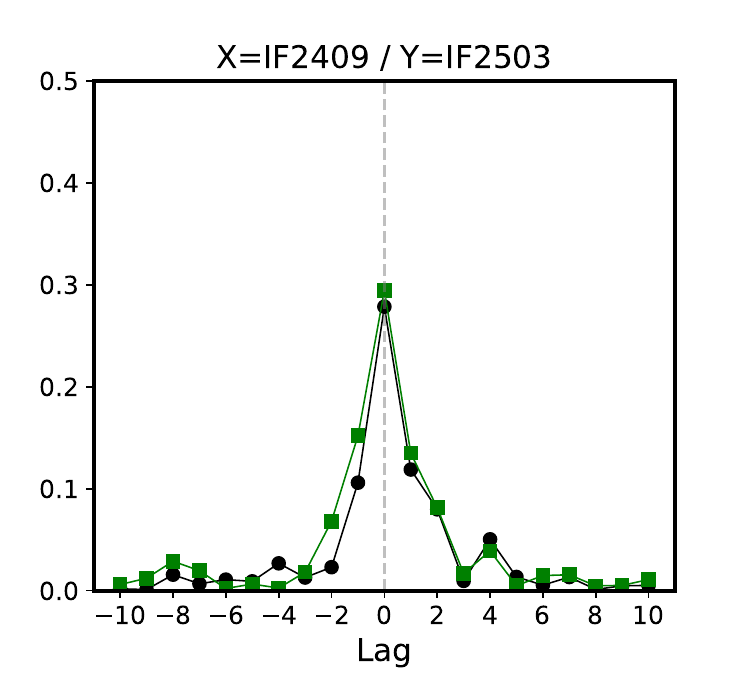}
	       \end{subfigure}
	       \begin{subfigure}{0.3\linewidth}
		  \includegraphics[width=\linewidth]{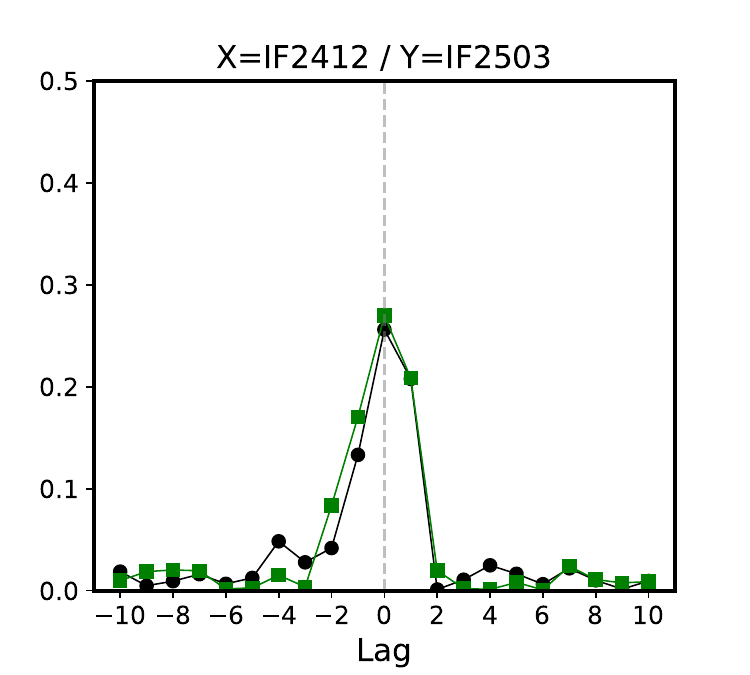}
	       \end{subfigure}
	\caption{The tick-time HY cross-correlation function of a sample dataset (e.g., November 24, 2020) of all pairs of futures contracts for the IF index. The most liquid futures contract is the near-month contract, IF2408. Based on the trading volume, the liquidity ranking for these contracts is as follows: IF2408 (near-month), IF2409 (next-month), IF2412 (quarterly), and IF2503 (semi-annual). The x-axis represents the `Lag', measured in time ticks, where one time tick corresponds to 500 milliseconds. The correlation value at a given lag $\ell$ represents the cross-correlation between $X$ and $Y$ with $Y$ shifted forward by $\ell$ time ticks, i.e., $\hat{\rho}_{HY}(\ell)$. In each panel, two cross-correlation values are shown: one calculated using the raw data (green line) and the other using transformed data (black line).}
	\label{fig: cross-correlation function}
\end{figure}

We calculate lead-lag correlation values on time-tick grids, ${-10, -9, \dots, 0, \dots, 9, 10}$, where each time-tick corresponds to 500 milliseconds. These calculations allow us to measure the lead-lag relationships between pairs of futures contracts with different maturities. We present example cross-correlation curves using a sample dataset (see Fig.~\ref{fig: cross-correlation function}) and compare the cross-correlation curves derived from raw data and transformed data. For details on the data transformation procedure and reasoning, refer to Section~\ref{sect: data}. 

We observe that the two cross-correlation curves align well in terms of both amplitude and shape. Therefore, the lead-lag statistics (e.g., LLT, LLR, LLC) extracted from the cross-correlation curves should be consistent. Based on this comparison, we conclude that the binning (i.e., event grouping) and filling (i.e., interpolation) processes applied to the original non-synchronous data do not significantly alter the lead-lag analysis outcomes. In Fig.~\ref{fig: cross-correlation function}, we find that the most liquid futures contract leads the other three contracts by one time-tick, with a very strong leading effect. The cross-correlation functions between $F_{1}$ and the other $F$ contracts exhibit an impulse-like shape, peaking at $\text{Lag} = 1$. Furthermore, we do not observe significant lead-lag relationships between other pairs of futures contracts. These findings are consistent across all trading days and all futures classes (i.e., IC, IF, IH, IM).

Table~\ref{tab: leadlag measures} presents summary statistics on the lead-lag measures for all pairs of futures contracts across the entire sample of trading days. The first three panels display the lead-lag measures for the lead futures contract ($F_{1}$) against the other three futures contracts. In terms of mean and median values, the LLTs and LLCs for pairs involving $F_{1}$ are similar, while the LLR increases with the liquidity ratio. For instance, the LLR for the pair $F_{1}$ vs. $F_{2}$ is lower than that for $F_{1}$ vs. $F_{3}$. The lead-lag relationships for pairs not involving $F_{1}$ are less pronounced. The sample quantiles, $Q{.05}$ and $Q{.95}$, show no significant or robust lead-lag relationships across the full sample. The last two columns report the one-sided statistical significance results from the bootstrap procedure described in Section~\ref{sect: method}. We focus on the first three panels, which include the pairs involving $F_{1}$. At the 5\% significance level, both LLR and LLT indicate that $F_{1}$ leads the other contracts by one tick on the majority of days.

\begin{table}[!htbp]
\caption{Descriptive statistics of pairwise lead-lag measures for IF} \label{tab: leadlag measures}
\centering
{
\def\sym#1{\ifmmode^{#1}\else\(^{#1}\)\fi}
\begin{tabular}{l*{1}{ccccccc}}
\toprule
                    &         Measures&     Mean&        $Q_{.05}$&          Median&         $Q_{.95}$&         $Q_{.05}^{LLR} < 1$&           $Q_{.05}^{LLT} < 1$\\
\midrule
$F_{1}$ vs. $F_{2}$ &        $LLT$          &      0.97&     1.00&     1.00&        1.00&      .&          32/700\\
                    &        $LLR$          &      5.23&      1.28&      4.45&          10.57&       30/700&          .\\
                    &        $LLC$          &      0.29&       0.2&        0.29&          0.37&       .&          .\\
$F_{1}$ vs. $F_{3}$ &        $LLT$          &      0.99&     1.00&     1.00&         1.00&      .&          8/700\\
                    &        $LLR$          &      8.47&      3.20&      7.46&          16.59&       2/700&          .\\
                    &        $LLC$          &       0.28&      0.18&        0.28&          0.36&          .&          .\\
$F_{1}$ vs. $F_{4}$ &        $LLT$          &      0.99&      1.00&     1.00&        1.00&      .&          16/700\\
                    &        $LLR$          &      10.78&      4.35&      9.76&          20.17&       2/700&          .\\
                    &        $LLC$          &       0.25&       0.15&        0.25&          0.34&          .&          .\\
$F_{2}$ vs. $F_{3}$ &        $LLT$          &      0.34&      0.00&     0.00&         1.00&      .&          599/700\\
                    &        $LLR$          &      2.58&      1.21&      2.38&          4.76&       77/700&          .\\
                    &        $LLC$          &      0.20&       0.12&        0.20&          0.30&      .&          .\\
$F_{2}$ vs. $F_{4}$ &        $LLT$          &      0.27&      0.00&     0.00&         1.00&      .&          639/700\\
                    &        $LLR$          &      3.34&      1.42&      2.95&          6.72&       69/700&          .\\
                    &        $LLC$          &      0.18&       0.10&        0.17&          0.26&      .&          .\\
$F_{3}$ vs. $F_{4}$ &        $LLT$          &      0.14&     0.00&     0.00&         1.00&      .&          671/700\\
                    &        $LLR$          &      1.89&      0.67&      1.56&          4.08&       355/700&          .\\
                    &        $LLC$          &       0.18&       0.09&        0.17&          0.28&          .&          .\\
\bottomrule
\end{tabular}
}

\small{Note: The lead-lag measures, LLT, LLR, and LLC, are presented for each pair of futures contracts in the IF dataset. LLTs are measured in tick-time, where one tick corresponds to approximately 500 milliseconds. The last two columns, $Q_{.05}^{LLR} < 1$ and $Q_{.05}^{LLT} < 1$, report the statistical significance of the lead-lag relationship, determined using a block bootstrap procedure. Specifically, $Q_{.05}^{LLR} < 1$, as shown in the first three panels of the table: $F_{1}$ vs. $F_{2}$, $F_{1}$ vs. $F_{3}$ and $F_{1}$ vs. $F_{4}$.
}
\end{table}

\begin{figure}[!ht]
      \centering
	   \begin{subfigure}{0.45\linewidth}
		\includegraphics[width=\linewidth]{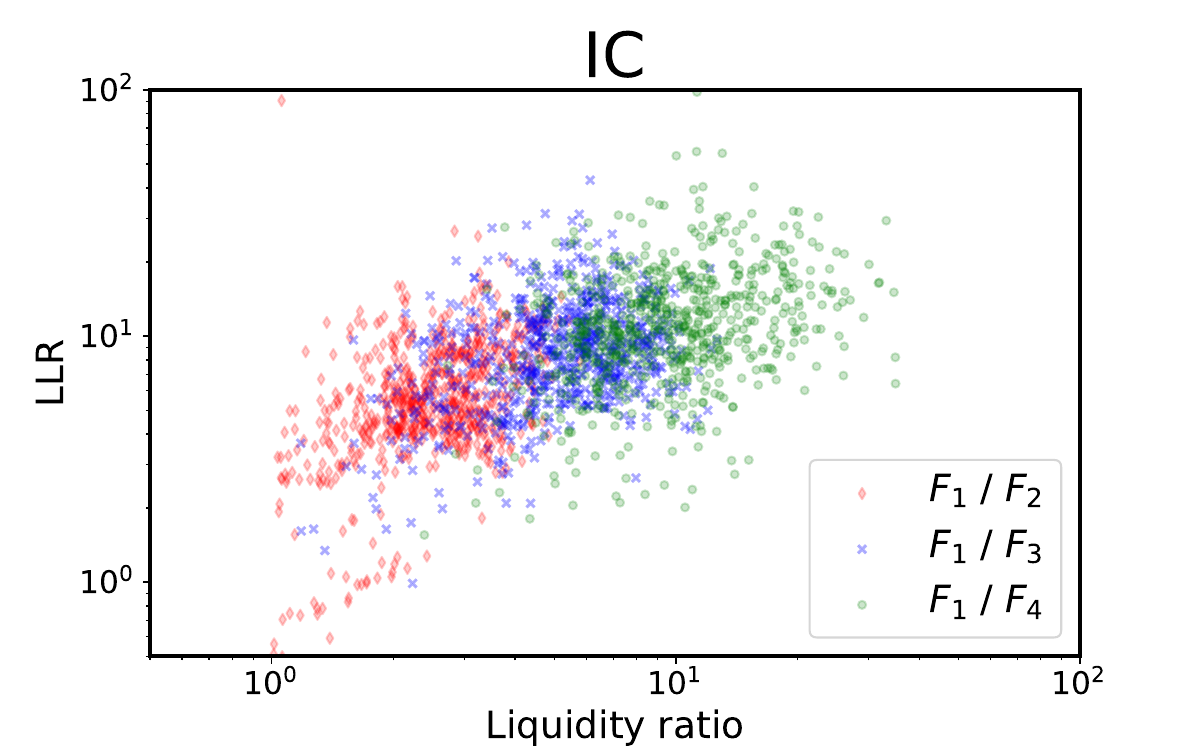}
		\label{fig:subfig1}
	   \end{subfigure}
	   \begin{subfigure}{0.45\linewidth}
		\includegraphics[width=\linewidth]{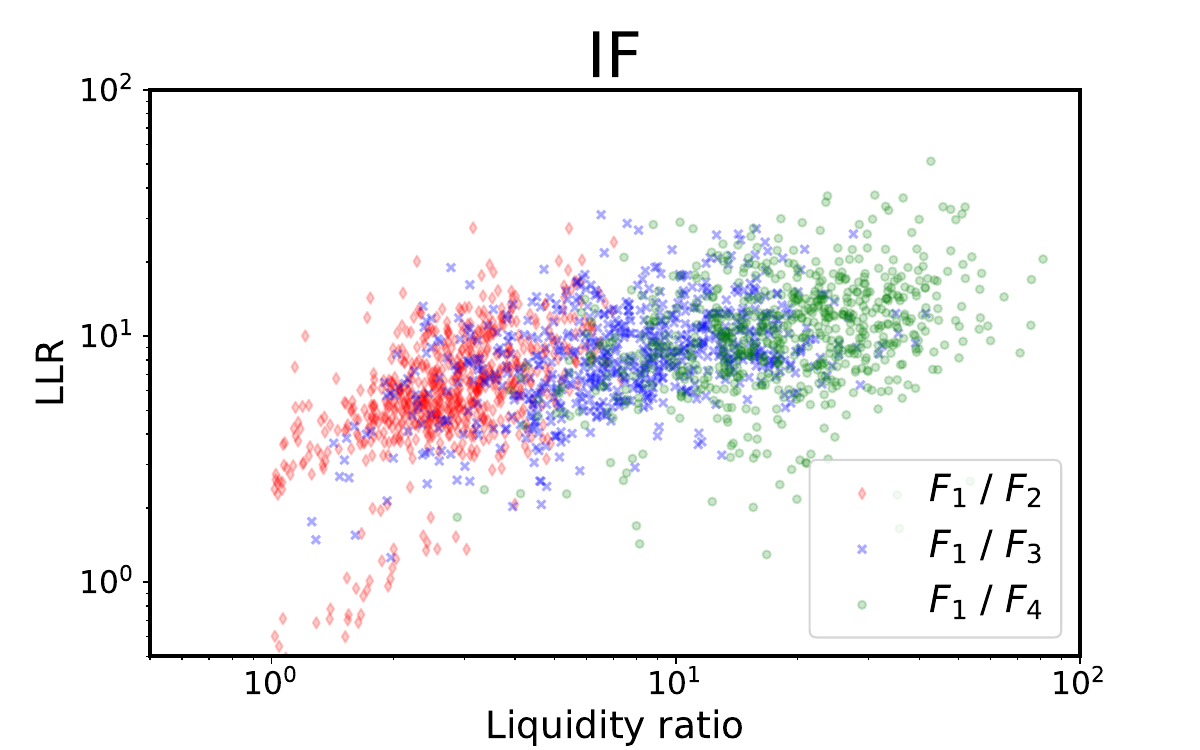}
		\label{fig:subfig2}
	    \end{subfigure}
	\vfill
	   \begin{subfigure}{0.45\linewidth}
		\includegraphics[width=\linewidth]{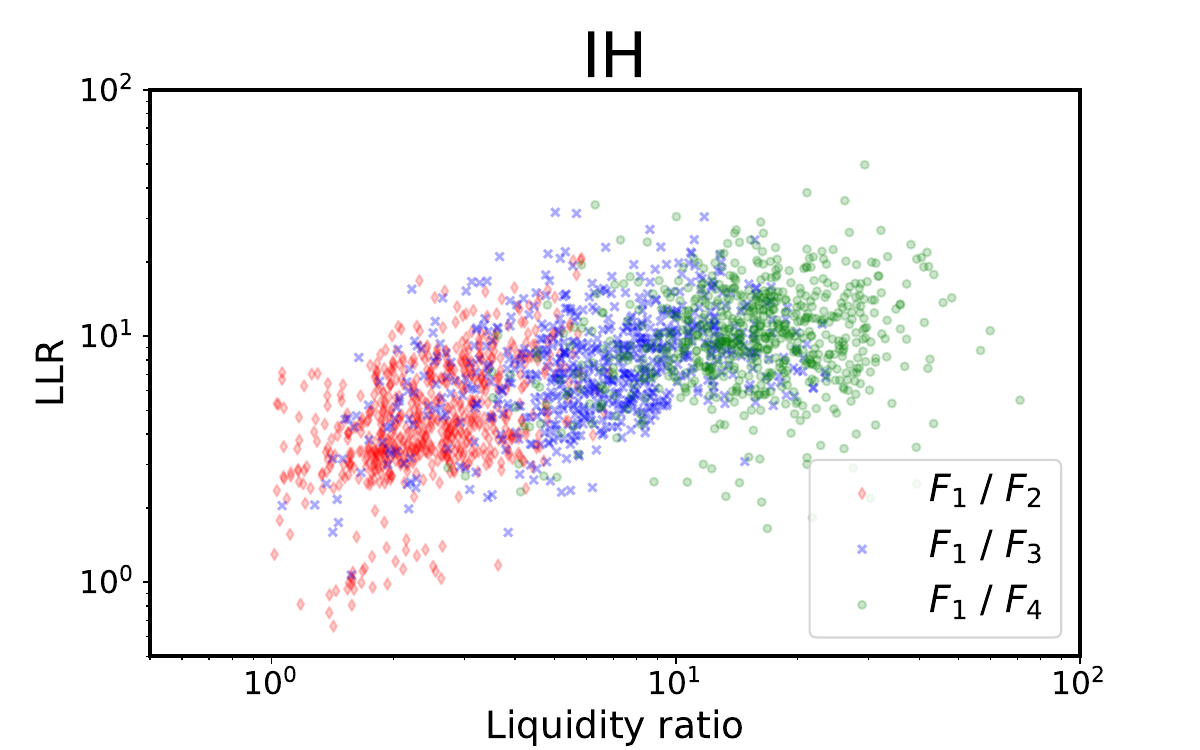}
		\label{fig:subfig1}
	   \end{subfigure}
	   \begin{subfigure}{0.45\linewidth}
		\includegraphics[width=\linewidth]{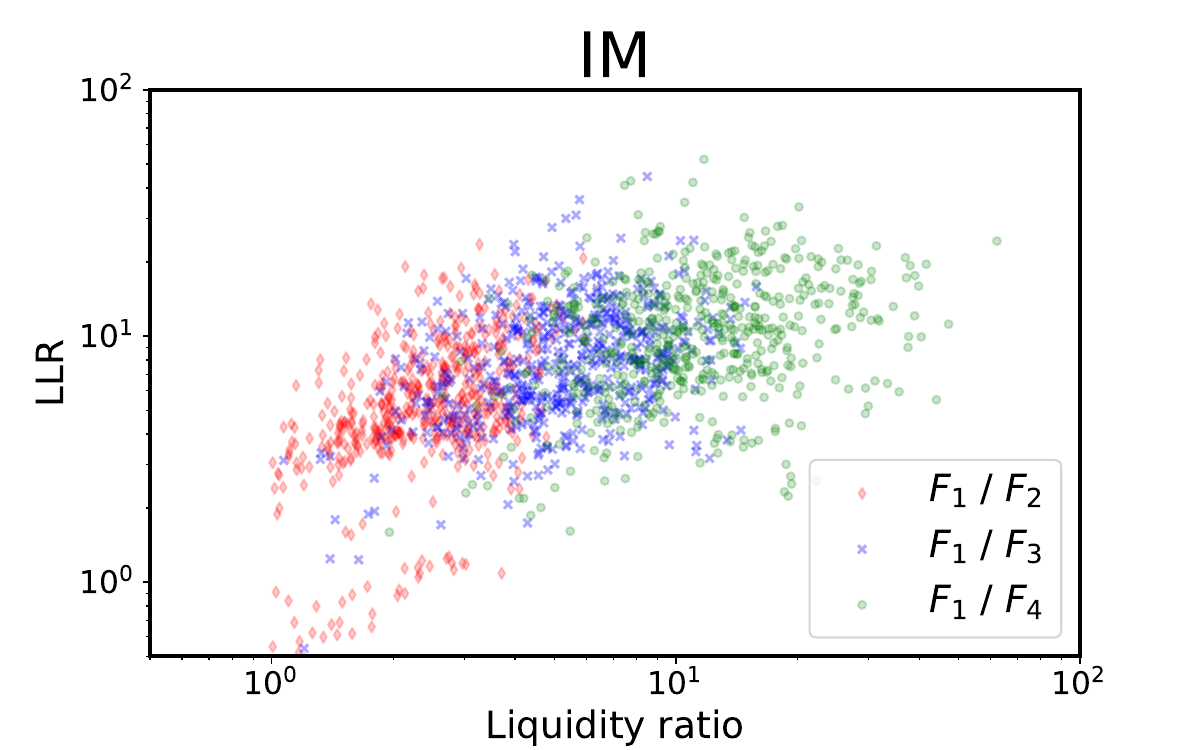}
		\label{fig:subfig2}
	    \end{subfigure}
	\caption{Scatterplot of LLR against relative liquidity ratio for all pairs of lead futures contract ($F_{1}$) and other futures contracts (i.e., $F_{2},\ F_{3},\ F_{4}$). The relative liquidity ratio is calculated as the ratio of trading volumes. In each panel, there are three pairs: $F_{1}$ vs. $F_{2}$ (red),  $F_{1}$ vs. $F_{3}$ (blue) and $F_{1}$ vs. $F_{4}$ (green).} 
	\label{fig: liquidity implied LLR}
\end{figure}

In Fig.~\ref{fig: liquidity implied LLR}, we plot the LLR against the relative liquidity ratio. Clearly, there is a positive relationship between LLR and liquidity ratio. In this study, we use trading volume as a measure of liquidity, where the liquidity ratio for a pair of futures contracts is defined as the ratio of their trading volumes. The results are consistent across all index futures: a higher liquidity ratio between two assets leads to a higher LLR for their lead-lag relationship. The choice of liquidity indicators (e.g., bid/ask spread, turnover, etc.) should not significantly affect the results, as these measures are highly correlated (\citealp{huth2014high}).

To quantify and extract the characteristic joint lead-lag network, we construct the minimum spanning tree (MST) from the lead-lag correlation matrix. The MST retains only the most significant correlations, and we find that the tree structure for most trading days is simple: {\resizebox{1.1em}{!}{\begin{tikzpicture}

\draw [line width=9pt](-3.5,4) node (v1) {} -- (-3.5,6.5);

\fill (-3.5,6.5) circle (15pt)  {};

\draw [line width=9pt](-3.5,4) -- (-5,6);
\fill (-5,6) circle (15pt)  {};
\draw [line width=9pt](-3.5,4) -- (-2,6);
\fill (-2,6) circle (15pt)  {};
\fill (-3.48,4) circle (15pt)  {};
\end{tikzpicture}}}, where the leading futures contract $F_{1}$ serves as the single root and leads the other contracts independently, see Table~\ref{tab: mst} for the proportion of days where the lead-lag network follows a "lead future-centered" structure.

\begin{table}[!htbp]
\caption{Descriptive statistics of the characteristic lead-lag network} \label{tab: mst}
\centering
{
\def\sym#1{\ifmmode^{#1}\else\(^{#1}\)\fi}
\begin{tabular}{l*{1}{ccccc}}
\toprule
                    &         IC&        IF&          IH&         IM\\
\midrule
Lead futures-centered tree &      662/700&     666/700&     612/700&         516/568\\
\bottomrule
\end{tabular}
}

\small{The numbers represent the proportion of days on which the minimum spanning tree (MST), extracted from the lead-lag correlation network, is structured such that the leading future is the root, and it leads the other futures contracts by one time tick.}
\end{table}

\subsection{Tick-by-tick term structure dynamics}\label{sec: term structure}

\begin{figure}[!ht]
      \centering
	   \begin{subfigure}{0.24\linewidth}
		\includegraphics[width=\linewidth]{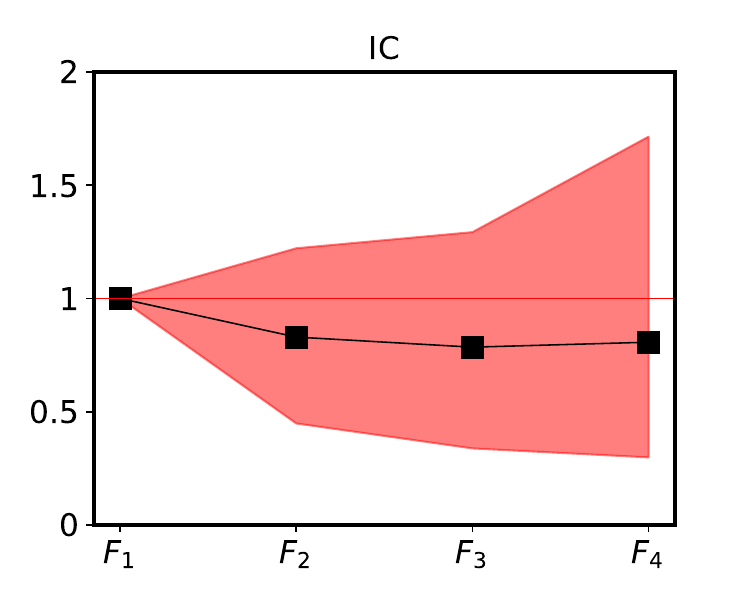}
	   \end{subfigure}
	   \begin{subfigure}{0.24\linewidth}
		\includegraphics[width=\linewidth]{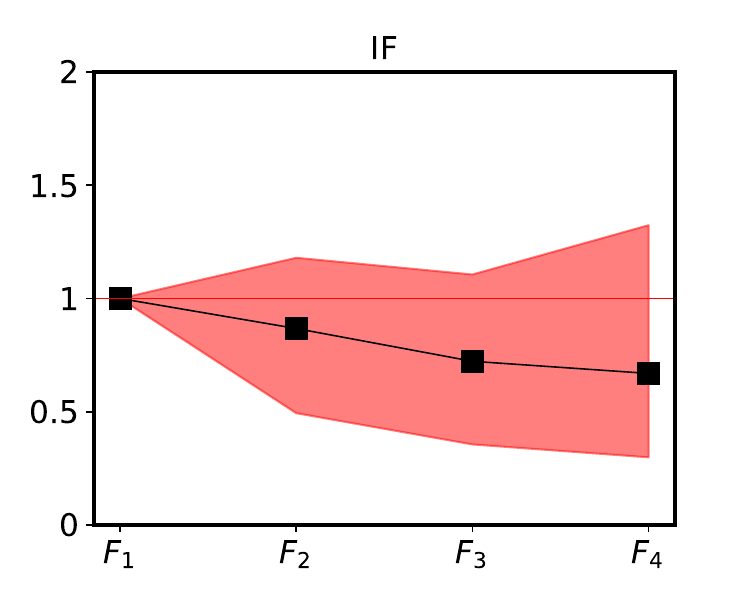}
	    \end{subfigure}
	   \begin{subfigure}{0.24\linewidth}
		\includegraphics[width=\linewidth]{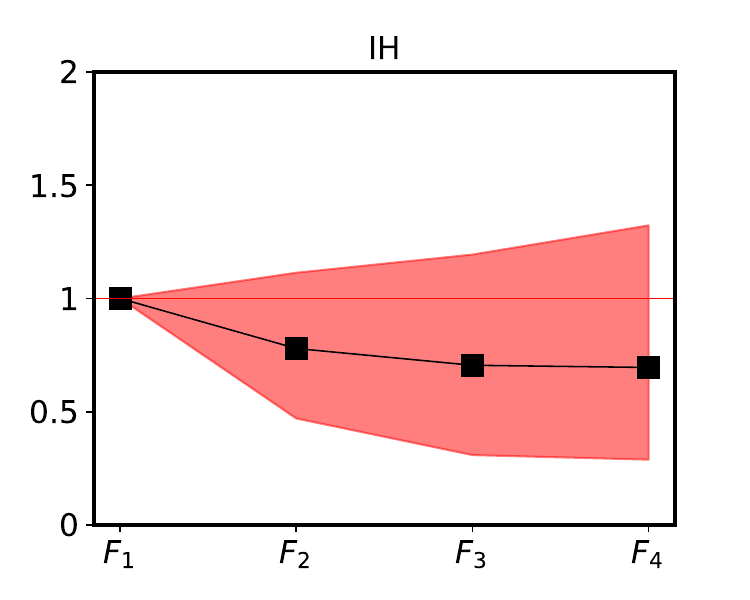}
	   \end{subfigure}
	   \begin{subfigure}{0.24\linewidth}
		\includegraphics[width=\linewidth]{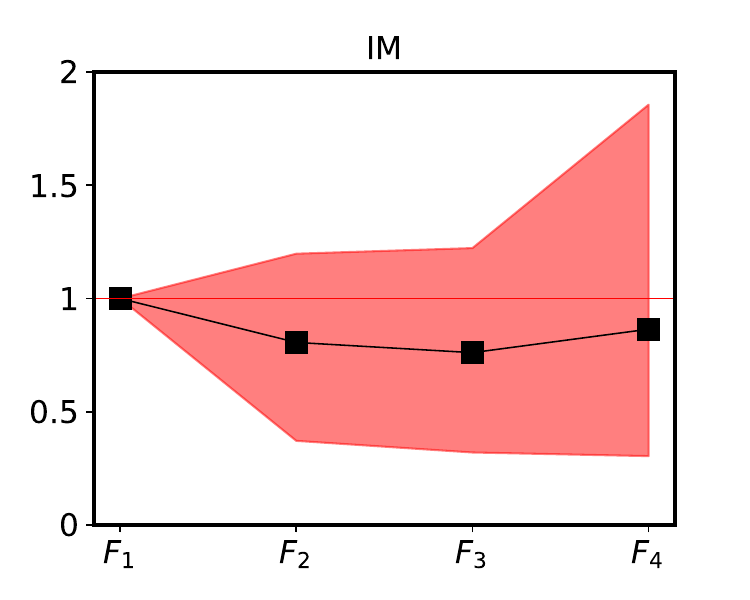}
	    \end{subfigure}
	\caption{First principal component of the joint variations in tick-by-tick changes of futures contracts across different maturities. This principal component, which explains approximately 50\% of the variance in tick-by-tick changes, can be interpreted as a level effect. The figures are normalized by rescaling the $F_{1}$ tick-change to 1, allowing for a clearer demonstration of the changes in other futures contracts that follow $F_{1}$'s movement.}
    \label{fig: tick-by-tick PC1}
\end{figure}

As the pairwise lead-lag analysis showed, the tick-by-tick shifts in the futures contracts are highly correlated across all maturities, in which lead futures contract $F_{1}$ moves first in temporal advance. In order to analyze the joint dynamics of all futures contracts, we apply PCA on the tick-by-tick variations data. We shift lead future's backward by one tick to eliminate the temporal influence from lead-lag relationship. The shape of first principal component is shown in Fig.~\ref{fig: tick-by-tick PC1}. All of its components are positive, thus it can be interpreted as a `level' factor. The shaded red area shows the 5\% to 95\% quantiles across the full sample, covering all trading days. Note that the 5\% boundary is still significantly above zero. After accounting for the lead behavior of $F_{1}$, we interpret the joint dynamics of futures contracts across different maturities as follows, after a jump in the lead futures contract, all other futures contracts are very likely to follow in the same direction, with a similar magnitude of change, though with some decay.

We do not present the shapes of the other principal components (e.g., $\text{PC}_{2}$) in this paper, as we find that their shapes, when computed on a daily basis, are not consistent across all trading days in the full sample. Additionally, the explained variance ratios for the second, third, and fourth principal components ($\text{PC}_{2}$, $\text{PC}_{3}$, and $\text{PC}_{4}$) are approximately 15\% each, whereas the first principal component ($\text{PC}_{1}$) accounts for about 50\% of the total variance, as shown in Table~\ref{tab: pca}. This suggests that the major joint movements in the data are captured by $\text{PC}_{1}$, with the contributions from the other components being relatively similar and not showing any dominant component.

Despite the results from pairwise and network lead-lag correlation analyses suggesting that the primary driver of tick-by-tick dynamics in futures contracts is the leading futures contract ($F_{1}$), the partial correlation between the lagged futures contracts ($F_{2}, F_{3}, F_{4}$), conditioned on changes in the leading futures contract, is still non-negligible. The conditional correlation can be calculated as follows (\citealp{kenett2015partial}):
\begin{align}\label{eq: partial corr}
\begin{split}
\rho(\Delta F_{i}(t), \Delta F_{j}(t)|\ \Delta F_{1}(t-1)) = \frac{\rho(\Delta F_{i}(t), \Delta F_{j}(t)) - \rho(\Delta F_{i}(t), \Delta F_{1}(t-1))\rho(\Delta F_{j}(t), \Delta F_{1}(t-1))}{\sqrt{(1-\rho^{2}(\Delta F_{i}(t), \Delta F_{1}(t-1)))^{2}(1-\rho^{2}(\Delta F_{j}(t), \Delta F_{1}(t-1)))^{2}}},
\end{split}
\end{align}
where $i\neq 1$, $j \neq 1$ and $i \neq j$. The pairwise correlation values are provided in the LLC data from Table~\ref{tab: leadlag measures}. By substituting correlation data into Eq.~\ref{eq: partial corr}, we can see that the conditional correlation is significantly greater than zero. While the "first-order" statistical analyses (e.g., lead-lag analysis, minimum spanning tree, principal component analysis) suggest a straightforward causal structure among futures contracts with different maturities, i.e., centered around the lead-future contract, a more comprehensive understanding of the statistical relationships among these contracts requires quantifying additional conditional dependencies. However, exploring these dependencies is beyond the scope of this paper.

\begin{table}[!htbp]
\caption{Descriptive statistics of the explained variance ratios for principal components of tick-by-tick price variations across all maturities of IF} \label{tab: pca}
\centering
{
\def\sym#1{\ifmmode^{#1}\else\(^{#1}\)\fi}
\begin{tabular}{l*{1}{cccc}}
\toprule
                    &        Mean&        $Q_{.05}$&          Median&         $Q_{.95}$\\
\midrule
$\text{PC}_{1}$ variance ratio  &        0.45          &      0.41&     0.46&     0.50\\
$\text{PC}_{2}$ variance ratio  &        0.21          &      0.18&     0.21&     0.25\\
$\text{PC}_{3}$ variance ratio  &        0.18          &      0.16&     0.18&     0.21\\
$\text{PC}_{4}$ variance ratio  &        0.16          &      0.13&     0.15&     0.17\\
\bottomrule
\end{tabular}
}

\small{Note: The numbers represent the daily variations in the explained variance ratios. Principal component analysis (PCA) is applied to tick-by-tick variations for each trading day, with the explained variance ratios summing to one for each data sample.}
\end{table}

\subsection{A negative feedback effect from calendar spread}\label{sec: regr results}

Given the strong short-term correlation between the leader-lagger pair, we aim to investigate whether the relationship between the leader and lagger can serve as a predictor for the leading asset’s return. We hypothesize that high-frequency traders (HFTs) could profit by trading the "lead-lag spread" when it deviates significantly. Specifically, when this spread diverges, it may impact the return of the leading asset, potentially creating a profitable trading signal. In this study, we use $F_{2}$ as the lagger asset to calculate the lead-lag spread. The leading futures contract  ($F_{1}$) is typically the near-month contract, while the lagger futures contract  ($F_{2}$) is often the next-month contract. The time difference between these contracts is short, which reduces the risk associated with the "repo" spread (which refers to the difference in financing costs for futures contracts with different tenors). For example, if the spread between $F_{1}$ and $F_{2}$, defined as $F_{1} - F_{2}$, increases sharply over a microscopic time scale (e.g., tick-by-tick), this often occurs when when market conditions are volatile. In such cases, HFTs might engage in a pair trade by selling $F_{1}$ and buying $F_{2}$ to capture this statistical arbitrage opportunity. Consequently, the price of $F_{1}$ could be affected. The lead-lag relationship has two sides. While much of the existing literature has focused on the potentially profitable opportunities arising from the lagging asset following the leader, our focus is on the opposite side: how the dynamic relationship between the leader and lagger can provide additional information about the leading asset itself.

Recall the two models introduced in Section~\ref{sect: method}: Model1 (Eq.~\ref{eq: model1}) and Model2 (Eq.~\ref{eq: model2}). These models test whether the short-term trend in the calendar spread exerts a negative feedback effect on the leading futures contract's return. In particular, we examine whether this predictive power is driven by the calendar spread itself, rather than by mean reversion in the individual futures contracts. Table~\ref{tab: regr} presents the regression results, which are divided into eight panels, one for each combination of forecasting horizon and model, with horizons $h = 1, 2, 4, 8$. In each panel, the first row shows the estimated parameters, and the second row displays the corresponding t-statistics. These estimated parameters and t-statistics are presented as intervals to capture daily variations, helping to assess the robustness and consistency of the results across all trading days in the dataset.

\begin{table}[!htbp]
\caption{Estimation results of Model1 and Model2 for IF} \label{tab: regr}
\centering
{
\def\sym#1{\ifmmode^{#1}\else\(^{#1}\)\fi}
\begin{tabular}{l*{1}{cccccc}}
\toprule
                    Forecasting horizon&         Model&  $\beta_{0}$&     $\beta_{\theta}$&        $\beta_{1}$&          $R^{2}$\\
\midrule
                    $h=1$ &                     Model1&    (-0.01, 0.01)&         (-0.21, -0.12)&     .&         (0.02, 0.07)\\
                     &                                &    \footnotesize{(-2.19, 2.10)}&    \footnotesize{(-49.18, -23.70)}&     .&         .\\
                     &                          Model2&     (-0.01, 0.01)&         (-0.22, -0.13)&     (0.00, 0.02)&         (0.02, 0.07)\\
                     &                                &    \footnotesize{(-1.82, 1.87)}&    \footnotesize{(-40.80, -24.50)}&     \footnotesize{(1.30, 11.20)}&         .\\
                    $h=2$ &                     Model1&    (-0.01, 0.01)&         (-0.24, -0.12)&     .&         (0.02, 0.05)\\
                     &                                &    \footnotesize{(-2.81, 2.76)}&    \footnotesize{(-37.02, -17.91)}&     .&         .\\
                     &                          Model2&    (-0.01, 0.01)&         (-0.25, -0.14)&     (0.00, 0.03)&         (0.02, 0.06)\\
                     &                                &    \footnotesize{(-2.52, 2.43)}&    \footnotesize{(-37,84, -19.45)}&     \footnotesize{(0.87, 15.22)}&         .\\
                    $h=4$ &                     Model1&    (-0.01, 0.01)&         (-0.29, -0.13)&     .&         (0.01, 0.04)\\
                     &                                &    \footnotesize{(-2.98, 2.77)}&    \footnotesize{(-31.80, -11.92)}&     .&         .\\
                     &                          Model2&    (-0.01, 0.01)&         (-0.30, -0.15)&     (-0.01, 0.04)&         (0.01, 0.05)\\
                     &                                &    \footnotesize{(-2.60, 2.55)}&    \footnotesize{(-32.50, -14.63)}&     \footnotesize{(-1.60, 17.40)}&         .\\
                    $h=8$ &                     Model1&    (-0.01, 0.01))&         (-0.34, -0.15)&     .&         (0.01, 0.03)\\
                     &                                &    \footnotesize{(-3.01, 3.00)}&    \footnotesize{(-25.93, -10.33)}&     .&         .\\
                     &                          Model2&    (-0.01, 0.01)&         (-0.35, -0.18)&     (-0.02, 0.02)&         (0.01, 0.04)\\
                     &                                &    \footnotesize{(-2.85, 2.79)}&    \footnotesize{(-26.38, -11.46)}&     \footnotesize{(-5.14, 4.14)}&         .\\
\bottomrule
\end{tabular}
}

\small{Note: For each forecasting horizon $h = 1, 2, 4, 8$ ticks, the coefficient estimates ($\beta$) and goodness of fit ($R^2$) are reported for both Model 1 and Model 2. For each scenario of forecasting horizon and model choice, the coefficient intervals are presented in the first line (in parentheses), and the $t$-statistics are reported in the second line (in parentheses). These intervals account for the daily variations in the estimates.}
\end{table}

\begin{figure}[!ht]
      \centering
	   \begin{subfigure}{0.48\linewidth}
		\includegraphics[width=\linewidth]{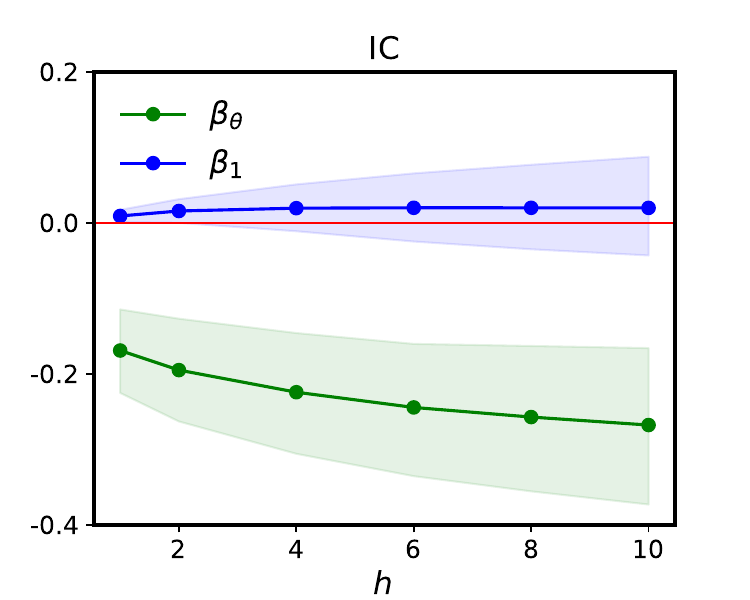}
	   \end{subfigure}
	   \begin{subfigure}{0.48\linewidth}
		\includegraphics[width=\linewidth]{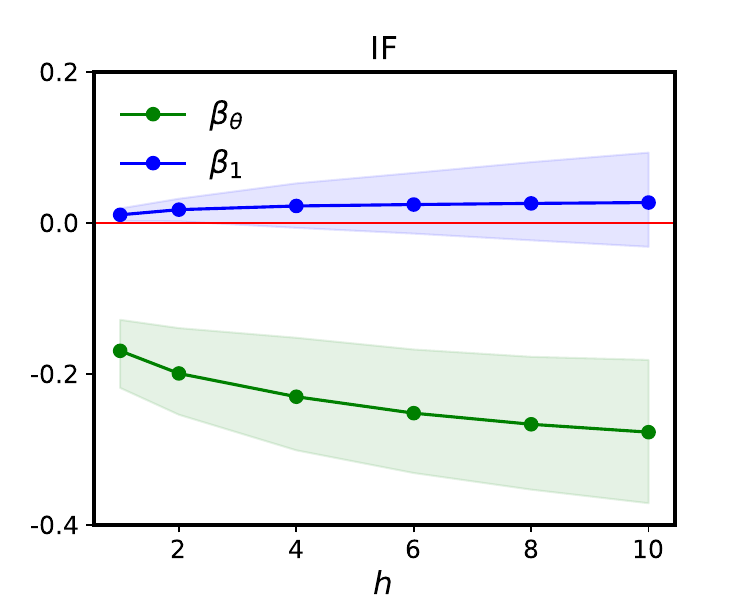}
	   \end{subfigure}
	\vfill
	   \begin{subfigure}{0.48\linewidth}
		\includegraphics[width=\linewidth]{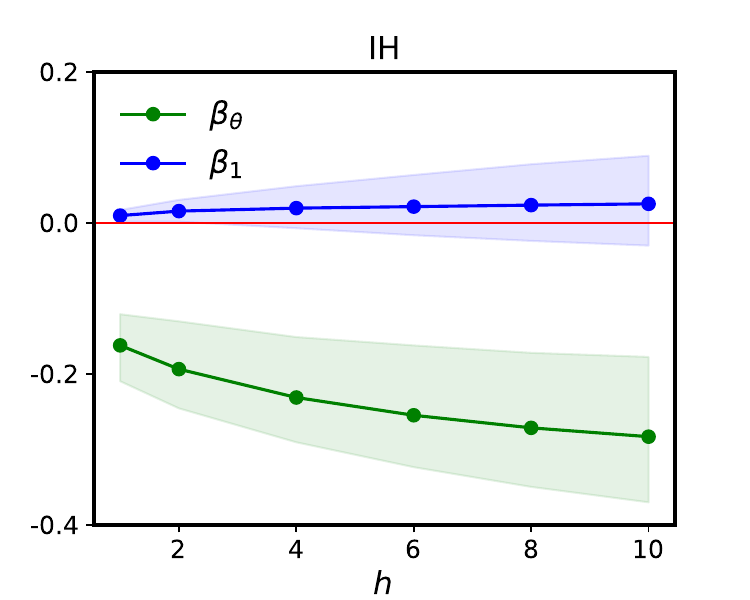}
	   \end{subfigure}
	   \begin{subfigure}{0.48\linewidth}
		\includegraphics[width=\linewidth]{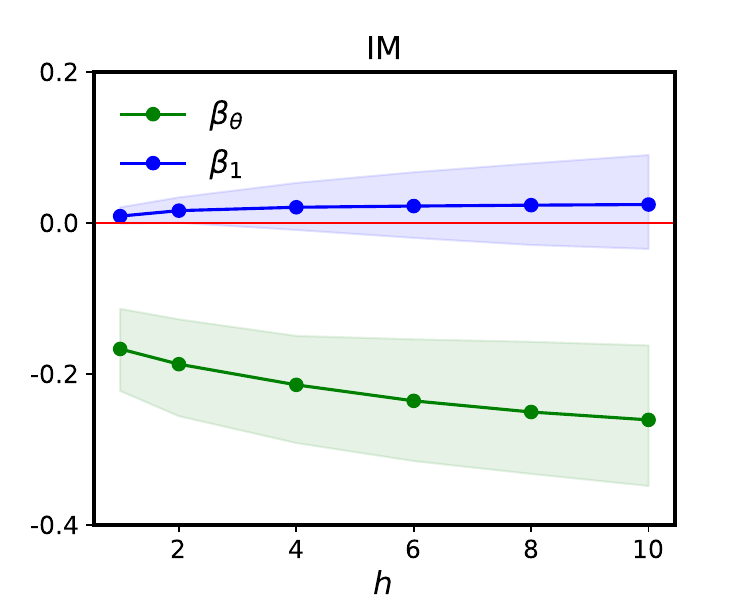}
	   \end{subfigure}
	\caption{Estimated coefficients from Model 2, which uses both short-term calendar spread momentum (coefficient $\beta_{\theta}$) and lead futures contract momentum (coefficient $\beta_{1}$). Ninety-five percent parameter intervals are represented by shaded areas: the green shaded area corresponds to $\beta_{\theta}$, and the blue shaded area corresponds to $\beta_{1}$. Note that these parameter intervals are not statistical confidence intervals derived from the linear regression procedure; rather, they account for the daily variations in the fitted parameters over the trading days included in the dataset. The dotted lines represent the average values across the trading days. The X-axis, $h$, represents the forecasting horizon.}
    \label{fig: regr coefs}
\end{figure}

From Table~\ref{tab: regr}, we observe that $\beta_{\theta}$ is significant in both Model1 and Model2, while $\beta_{1}$ in Model2 is notably smaller. Despite the small estimated value of $\beta_{\theta}$, the t-statistics remain relatively large, suggesting statistical significance for most trading days. In Fig.~\ref{fig: regr coefs}, we see that the parameter intervals for $\beta_{1}$ are wide and include zero, indicating that its directional effect varies on a daily basis. In contract, the calendar spread ($\beta_{\theta}$) consistently acts as a negative feedback on the return of the leading futures contract.

\begin{figure}[!ht]
      \centering
	   \begin{subfigure}{0.45\linewidth}
		\includegraphics[width=\linewidth]{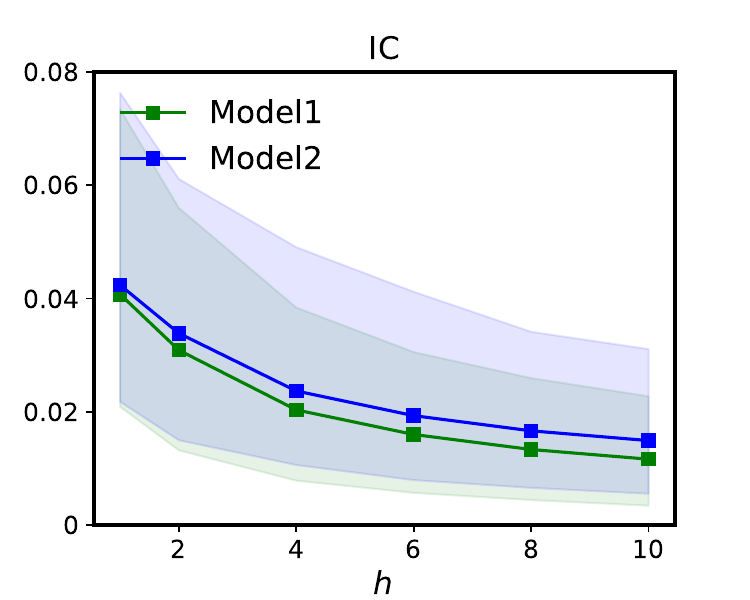}
	   \end{subfigure}
	   \begin{subfigure}{0.45\linewidth}
		\includegraphics[width=\linewidth]{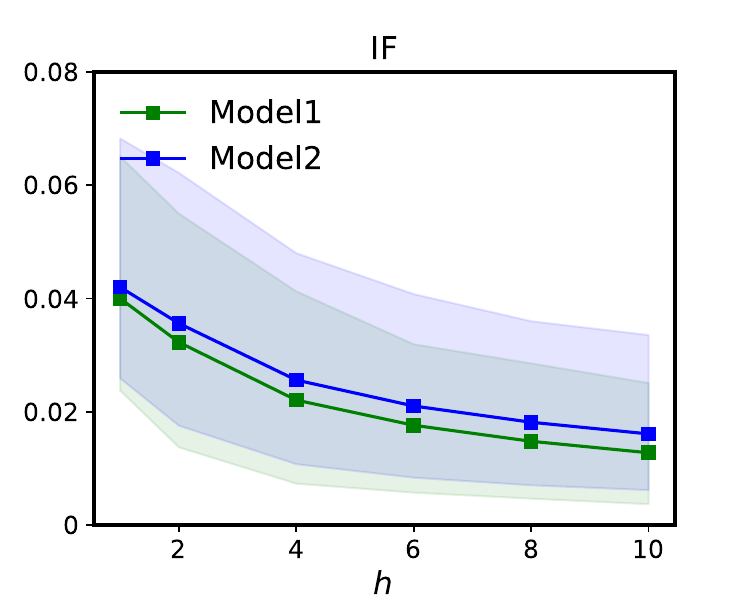}
	    \end{subfigure}
	\vfill
	   \begin{subfigure}{0.45\linewidth}
		\includegraphics[width=\linewidth]{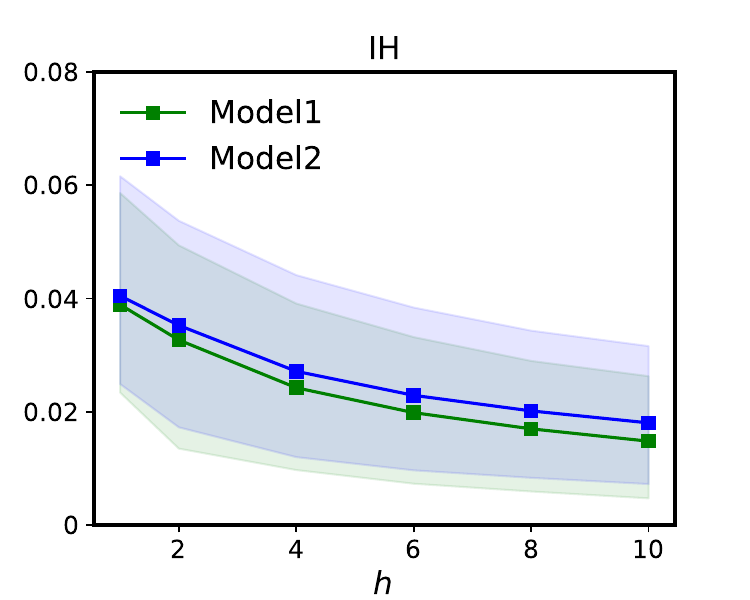}
	   \end{subfigure}
	   \begin{subfigure}{0.45\linewidth}
		\includegraphics[width=\linewidth]{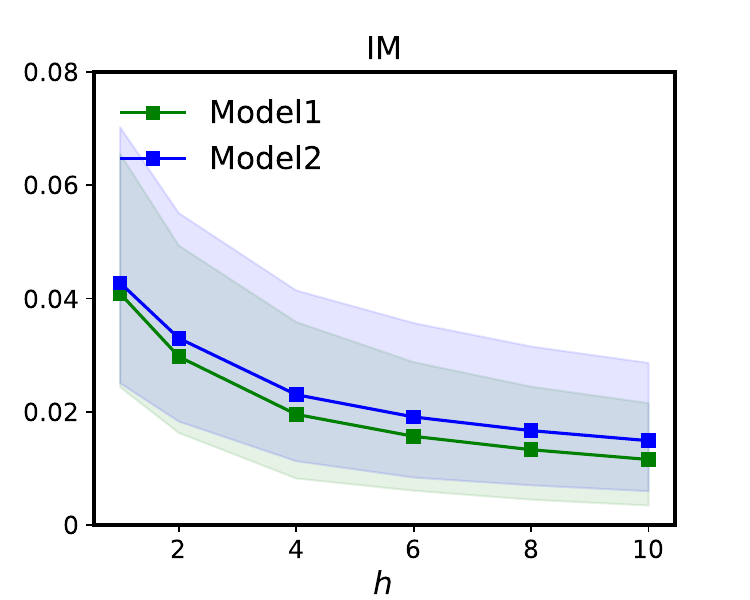}
	    \end{subfigure}
	\caption{The $R^2$ values for Model 1 and Model 2 vary with forecasting horizons ($h$). Ninety-five percent $R^2$ intervals are indicated by shaded areas: the green shaded area represents Model 1, and the blue shaded area represents Model 2. These $R^2$ intervals account for the daily variations in the fitted parameters over the trading days included in the dataset. The dotted lines represent the average values across the trading days.}
    \label{fig: regr R2}
\end{figure}

As shown in Fig.~\ref{fig: regr R2}, the improvement in $R^{2}$ from Model1 to Model2 is minimal. The forecasting accuracy, measured by $R^{2}$, deteriorates as the forecasting horizon increases, which is reflected in the widening distribution of returns. Based on these findings, we conclude that the predictive power of the lead futures contract's returns is not driven by mean reversion in the individual contracts. Instead, the primary driver is the negative feedback effect of the calendar spread. This conclusion holds consistently across all futures contracts traded on CFFEX, including IC, IF, IH, and IM, as shown in Fig.~\ref{fig: regr coefs} and Fig.~\ref{fig: regr R2}.

\subsection{Backtest performance}\label{sec: backtest}
As discussed in Section~\ref{sec: regr results}, the short-term calendar spread trend is a statistically significant signal for predicting the returns of the leading futures contract. This signal helps us estimate both the direction and magnitude of the mid-quote change over a short-term time horizon. However, rather than explicitly forecasting the returns of the leading futures contract, we focus on calibrating the strategy by optimizing the signal threshold that triggers trades. Specifically, we introduce a strategy parameter, the signal threshold ($\lambda$). When the short-term calendar spread trend ($\theta$), i.e., the signal, exceeds the threshold ($|\theta| \geq \lambda$), we initiate a trade on the leading futures contract at the current bid/ask prices.

\begin{figure}[!ht]
      \centering
	   \begin{subfigure}{0.45\linewidth}
		\includegraphics[width=\linewidth]{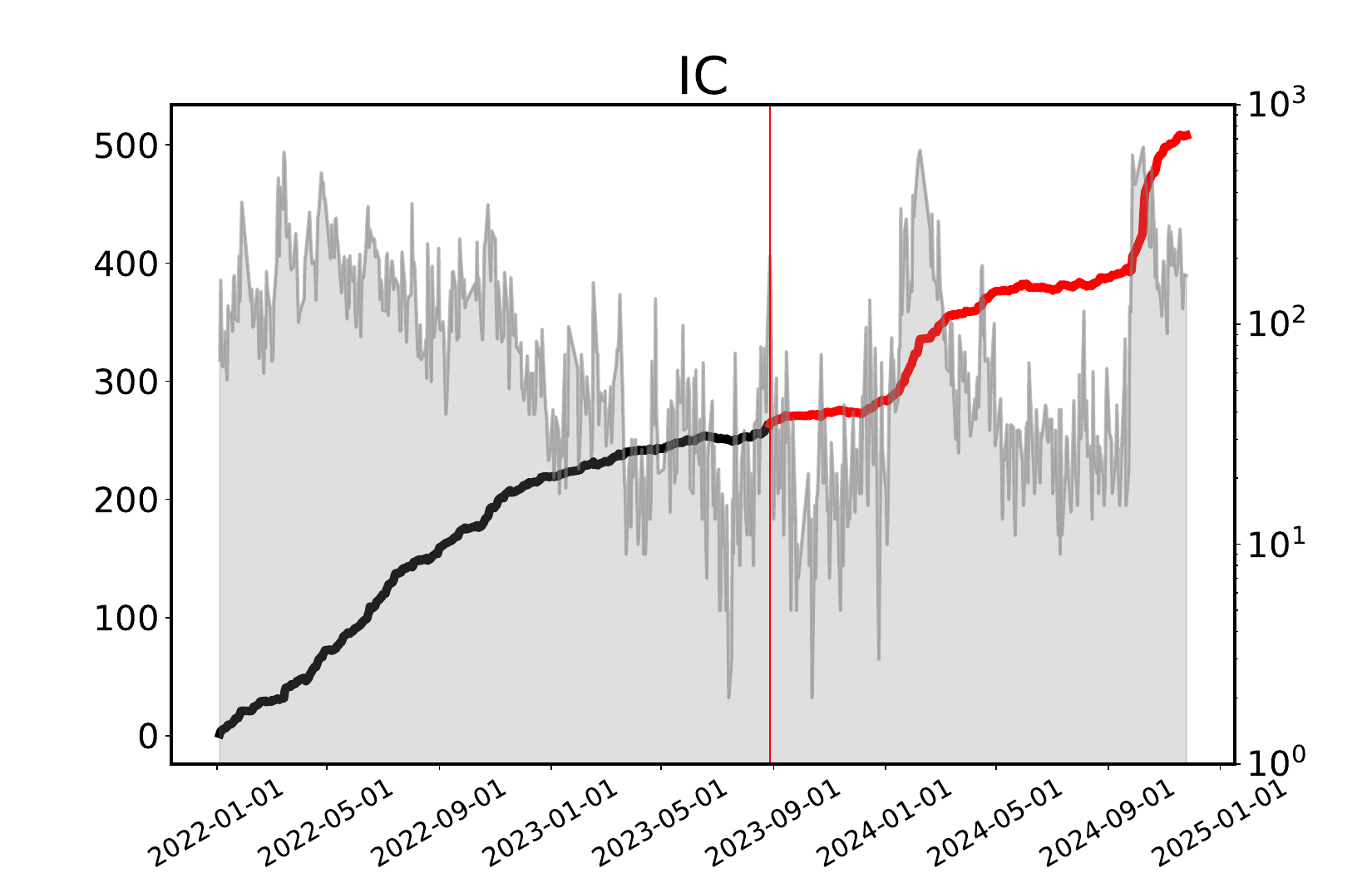}
	   \end{subfigure}
	   \begin{subfigure}{0.45\linewidth}
		\includegraphics[width=\linewidth]{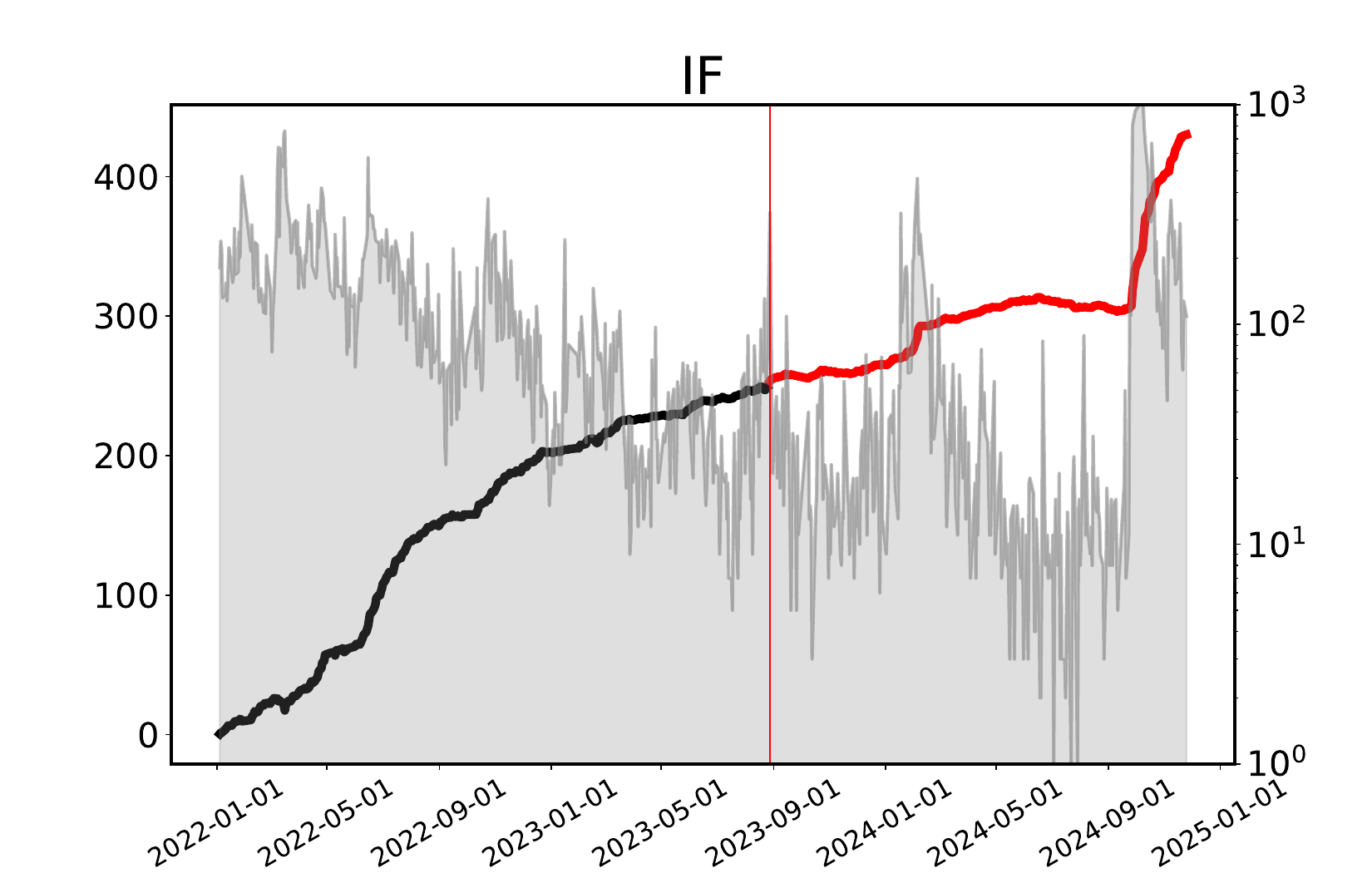}
	    \end{subfigure}
	\vfill
	   \begin{subfigure}{0.45\linewidth}
		\includegraphics[width=\linewidth]{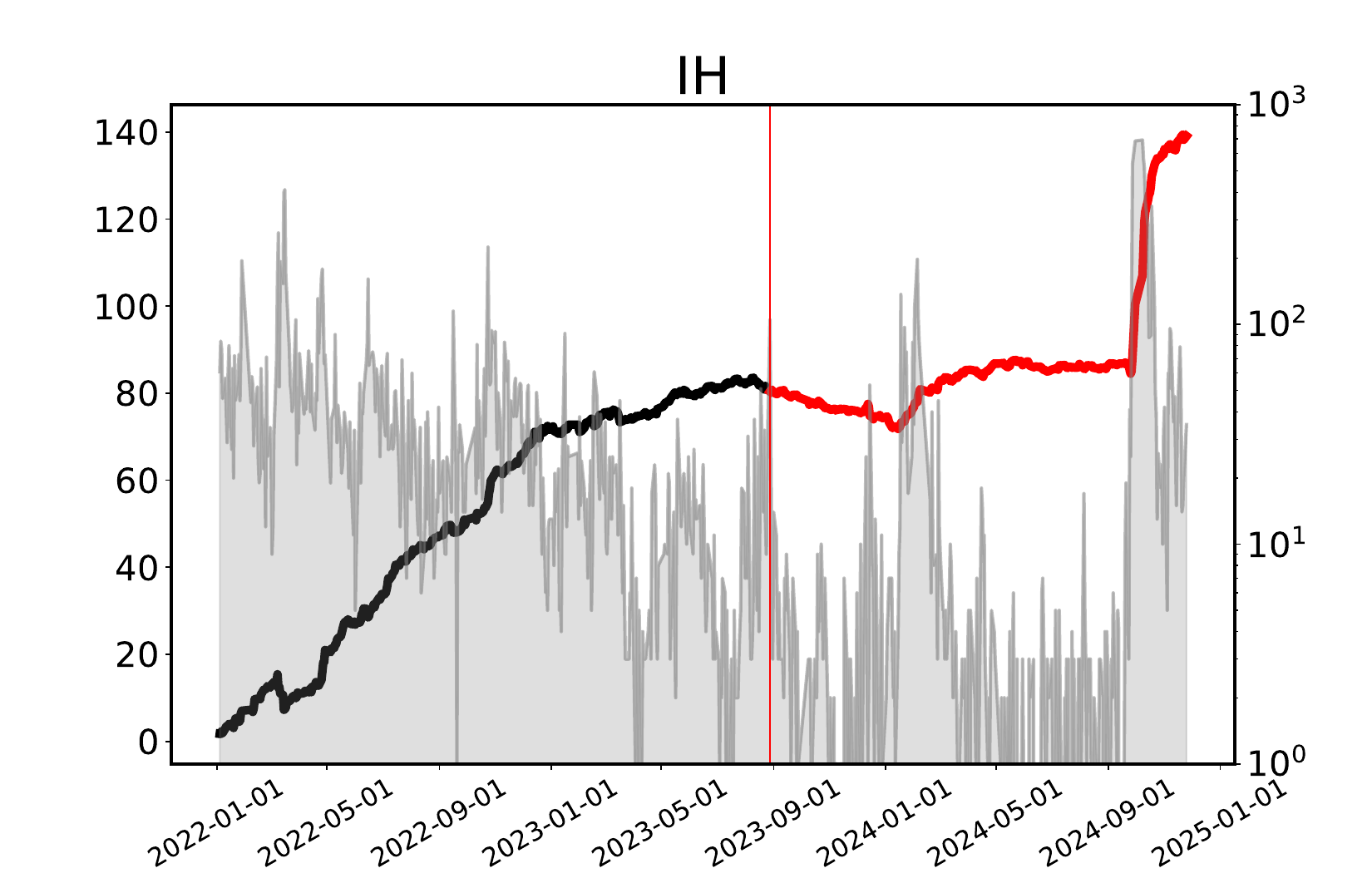}
	   \end{subfigure}
	   \begin{subfigure}{0.45\linewidth}
		\includegraphics[width=\linewidth]{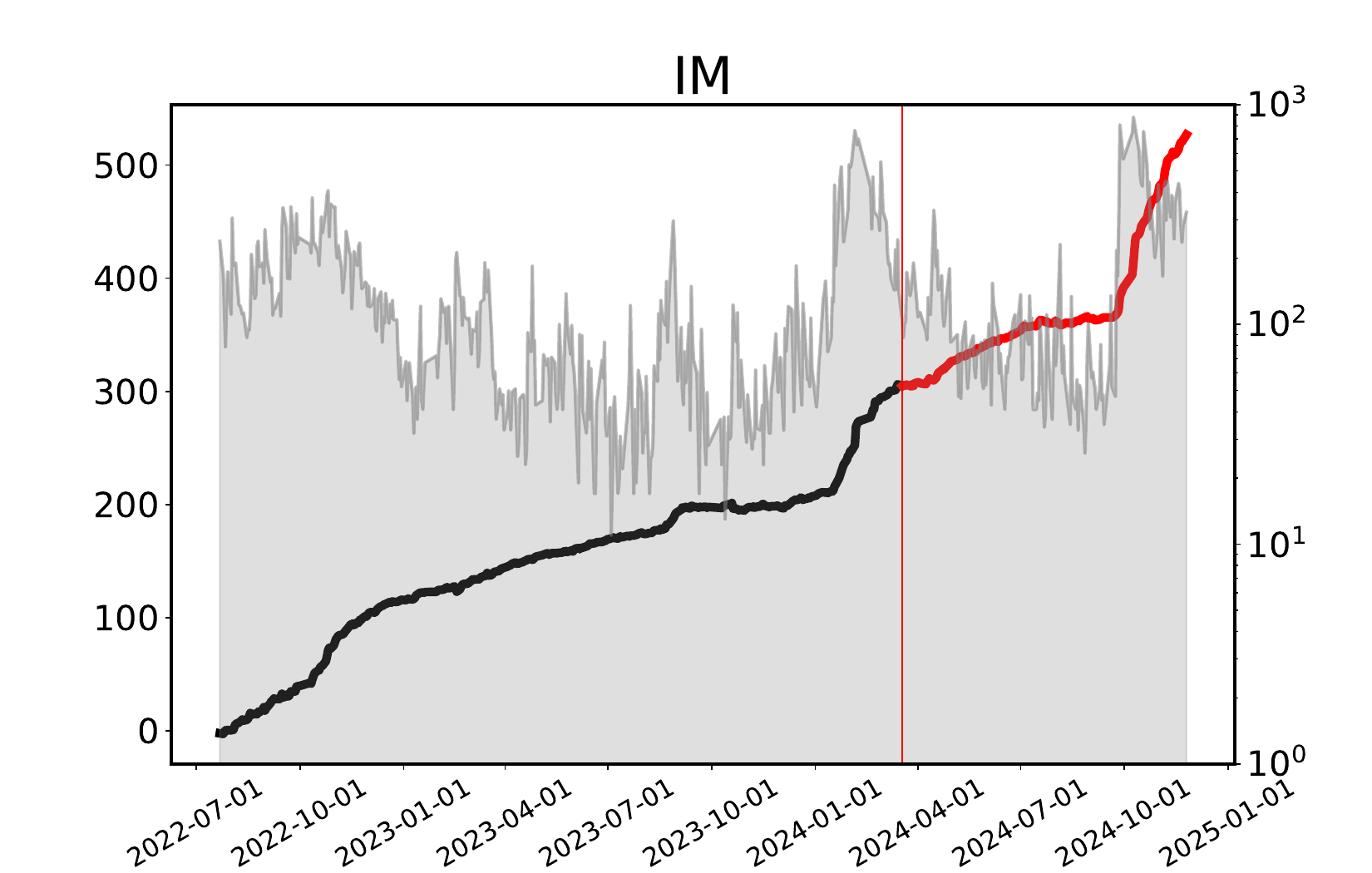}
	    \end{subfigure}
	\caption{Cumulative profit and loss (PnL) of the calendar spread feedback strategy on a daily basis. For each panel, the left Y-axis represents the cumulative PnL (in units of 10,000 Yuan), while the right Y-axis represents the number of trades per day. Each panel is divided into two regimes, in-sample performance (left, with PnL curve is in black) and out-of-sample performance (right, with PnL curve is in red).}
    \label{fig: backtest}
\end{figure}

For example, consider a strategy with a threshold parameter of $\lambda = 4$ price ticks. The minimum price tick is 0.2 Yuan for all futures contracts traded on CFFEX. At a given timestamp, if the signal value is $\theta = 5$ price ticks, we predict that the price of the leading futures contract will decrease. Consequently, we sell one unit of the leading futures contract at the market bid price. To minimize the bid/ask spread cost, trades are only executed when the Level 1 bid/ask spread of the leading futures contract is no greater than two price ticks. Positions are closed at the end of each trading day, and the maximum position size is capped at one lot. For each trade, we apply a transaction fee, which, based on CFFEX guidelines, is approximately half of one price tick. The profit and loss (PnL) curves are shown in Fig.~\ref{fig: backtest}

In Table~\ref{tab: backtest}, we present the backtest results for both in-sample and out-of-sample periods. The first 400 trading days are used as the in-sample data, and we determine the parameter that maximizes the Sharpe ratio. We then test the strategy using the out-of-sample data. There is no significant qualitative difference in trading performance between the in-sample and out-of-sample periods; the parameters remain stable across the entire sample. When comparing different index futures, we observe that the IC and IM contracts perform better in terms of PnL per trade and Sharpe ratio. This superior performance is likely due to the higher volatility of these contracts, which makes them more responsive to short-term market trends and events, potentially leading to more trading opportunities. Moreover, as shown in Fig.~\ref{fig: backtest}, we observe two significant jumps in the PnL curve (as well as the number of trades) for all stock index futures in 2024. The first jump occurs at the end of January and the beginning of February, while the second jump takes place at the end of September and the beginning of October. Both periods correspond to extreme market conditions, triggered by a series of positive policies released to stimulate the market. During these periods, stock index volatilities exceed 80\% for all indexes, and all major indexes experience daily jumps of 5\%. After accounting for bid/ask spread costs and transaction fees, our strategy consistently generates a profit of more than one price tick per trade in the out-of-sample test, a result that holds across all futures contracts.

\begin{table}[!htbp]
\caption{Backtest performance of calendar spread feedback strategy}
\centering
{
\def\sym#1{\ifmmode^{#1}\else\(^{#1}\)\fi}
\begin{tabular}{l*{1}{cccccc}}
\toprule
                     &                        Measures&        IC&        IF&        IH&          IM\\
\midrule
                     Parameter&                Signal threshold ($\lambda$)&         1.20&         0.80&        0.80&         1.20\\
                     In-sample&                 Number of trade per day&      110.34&        121.81&      34.95&         127.49\\
                     &                          PnL per trade &            0.30&         0.17&        0.19&         0.30\\
                     &                          Sharpe ratio &         9.68&         9.16&        4.86&         7.90\\
                     Out-of-sample&                Number of trade per day&      87.54&        71.87&        25.96&         167.66\\
                     &                          PnL per trade &            0.47&         0.28&        0.25&         0.40\\
                     &                          Sharpe ratio &         6.60&         5.26&        2.84&         8.79\\
\bottomrule
\end{tabular}
}

\small{Note: The dataset is split into two parts: in-sample data, used for calibrating the strategy parameters, and out-of-sample data, used to test the strategy's performance on unseen data. Performance is evaluated using several metrics, including the average PnL per day, average number of trades per day, average PnL per trade, and the annualized Sharpe ratio, with each metric averaged across the trading days in its respective group.}
\label{tab: backtest}
\end{table}

It is important to note that we do not account for execution risk in the backtest, which can be complex and subject to various uncertainties. For example, our decisions are based on signals derived from market data, specifically a stream of limit order books. As explained in Section~\ref{sect: data}, we calculate the signal when we receive both $F_{1}$ and $F_{2}$ order books. Although the time gap between receiving $F_{1}$ and $F_{2}$ is minimal (e.g., less than 0.01 milliseconds), there is still a possibility of missing a trading opportunity on $F_{1}$ due to delays in sending orders. Additionally, execution risk may lead to losing trades when no competition exists, or missed opportunities when many competitors are vying for the same trade, potentially introducing selection bias.

\section{Conclusion}\label{sect: conclusion}
This study offers a detailed examination of the lead-lag relationships among stock index futures contracts of varying maturities on the China Financial Futures Exchange (CFFEX). Through the application of advanced statistical techniques, including the Hayashi–Yoshida (HY) cross-correlation estimator and principal component analysis, we have confirmed that the most liquid futures contract consistently leads the others. This finding validates the hypothesis that liquidity plays a crucial role in price discovery, especially in high-frequency markets. Our results reveal a strong interconnectedness between futures contracts with different maturities, with the movements of the lead futures contract driving price adjustments in the others. This demonstrates that when the lead futures contract experiences price changes, the other contracts tend to follow in close succession, highlighting the importance of liquidity in shaping these dynamics. Additionally, we identify a negative feedback effect from the calendar spread on the leading futures contract, suggesting that this spread could serve as an effective predictor of future price movements in the lead contract.

The insights gained from this study have significant implications for both academic research and practical trading strategies. Future research could extend this work by exploring cross-product lead-lag relationships, such as between different types of stock index, to further characterize the temporal network of futures market. Moreover, our identification of the negative feedback effect opens the door for incorporating calendar spreads as an additional feature to forecast leading futures contract, offering traders a valuable tool for anticipating price movements. Altogether, this paper contributes to the growing body of literature on futures market microstructure, offering new insights into the role of liquidity in price discovery and providing actionable strategies for market participants seeking to capitalize on lead-lag relationships in high-frequency trading environments.

\subsection*{Acknowledgments}
The authors would like to express their gratitude to the members of the Market Making team at Equity Derivatives Trading for their insightful discussions and valuable contributions to this research.

\clearpage
\bibliographystyle{plainnat}
\bibliography{references}
\end{document}